\documentclass[%
superscriptaddress,
frontmatterverbose, 
showkeys,
nofootinbib,
nobibnotes,
amsmath,amssymb,
]{revtex4-2}

\pdfoutput=1
 
\catcode`\@=11
\def\lsim{\mathrel{\mathpalette\@versim<}}
\def\gsim{\mathrel{\mathpalette\@versim>}}
\def\@versim#1#2{\vcenter{\offinterlineskip
		\ialign{$\m@th#1\hfil##\hfil$\crcr#2\crcr\sim\crcr } }}
\catcode`\@=12

\catcode`@=12
\usepackage[export]{adjustbox}	
\usepackage{float}
\usepackage{epsfig,bbm}
\usepackage{amsmath}
\usepackage{slashed}
\usepackage{tikz}
\usepackage{bm}
\usepackage{amssymb}
\usepackage{mathtools}
\usepackage[caption=false]{subfig}
\usepackage{graphicx}
\usepackage{hyperref}
\usepackage{cleveref}% add hypertext capabilities
\usepackage{tabularx}
\usepackage{multirow}
\usepackage{stackengine}

\usepackage{lineno}

\def\BF{{\cal B}}

\def\lsim{\mathrel{\rlap{\lower4pt\hbox{\hskip1pt$\sim$}}
		\raise1pt\hbox{$<$}}} %less than or approx. symbol
\def\gsim{\mathrel{\rlap{\lower4pt\hbox{\hskip1pt$\sim$}}
		\raise1pt\hbox{$>$}}} %greater than or approx. symbol

\definecolor{sRed}{rgb}{0.6,0.1,0.1} 

\newcommand{\scp}[1]{{\color{black} #1}}

\newcommand{\yjk}[1]{{\color{black} #1}}

\begin{document}
%\preprint{APS/123-QED}

\title{ A comprehensive study of vector leptoquark \scp{with $U(1)_{B_3-L_2}$} \\ on the $B$-meson and Muon g-2 anomalies}

\author{Kayoung Ban}
\email{ban94gy@yonsei.ac.kr}
\affiliation{Department of Physics and IPAP, Yonsei University, Seoul 03722, Republic of Korea }

\author{Yongsoo Jho}
\email{yongsoo.jho@weizmann.ac.il}
%\affiliation{Department of Physics and IPAP, Yonsei University, Seoul 03722, Republic of Korea }
\affiliation{Weizmann Institute, Department of Particle Physics and Astrophysics, Rehovot, Israel 7610001}

\author{Youngjoon Kwon}
\email{yjkwon63@yonsei.ac.kr }
\affiliation{Department of Physics and IPAP, Yonsei University, Seoul 03722, Republic of Korea }

\author{Seong Chan Park}
\email{sc.park@yonsei.ac.kr}
\affiliation{Department of Physics and IPAP, Yonsei University, Seoul 03722, Republic of Korea }
%\affiliation{Korea Institute for Advanced Study, Seoul 02455, Republic of Korea}

\author{Seokhee Park}
\email{seokhee.park@kek.jp}
\affiliation{High Energy Accelerator Research Organization (KEK), Tsukuba 305-0801, Japan}

\author{Po-Yan Tseng}
\email{pytseng@phys.nthu.edu.tw}
\affiliation{Department of Physics, National Tsing Hua University, Hsinchu 300, Taiwan}

%%%%%%%%%%%%%%%%%%%%%%%%
\begin{abstract}
	Recently reported anomalies in various $B$ meson decays and also in the anomalous magnetic moment of muon $(g-2)_\mu$ motivate us to consider a particular extension of the standard model
	incorporating new interactions in lepton and quark sectors simultaneously. Our minimal choice would be leptoquark. In particular, we take vector leptoquark ($U_1$) and comprehensively study all related observables including ${(g-2)_{\mu}},\ R_{K^{(*)}},\ R_{D^{(*)}}$, $B \to (K) \ell \ell' $ where $\ell\ell'$ are various combinations of $\mu$ and $\tau$, and also lepton flavor violation in the $\tau$ decays. 
{We find that a hybrid scenario with additional $U(1)_{B_3-L_2}$ gauge boson provides a common explanation 
of all these anomalies.}
\end{abstract}
%%%%%%%%%%%%%%%%%%%%%%%%

\maketitle

	%%%%%%%%%%%%%%%%%%%%%%%%
	%%%%%%%%%%%%%%%%%%%%%%%%
	\section{Introduction}
	
	Physics is a data-driven science, and we are keen to modify the standard model (SM) when experimental results  deviate from the theoretical predictions. Over the past several years, the $B$-physics experiments BaBar, Belle, and LHCb have reported several anomalous results in the $b \to s \ell \ell$ and $b \to c \ell \nu$ processes, \scp{which are not properly explained within the SM thus call for new physics.}
	In particular, the lepton flavor universality (LFU), which is one of the approximate symmetries in the SM, seems to be broken beyond the expected range
	according to the observables of $R_D$, $R_{D^*}$, $R_K$, and $R_{K^*}$,
	which measured the ratios of different lepton flavors 
	\begin{align}
	R_{D^{(*)}}\equiv \frac{\BF(B \to D^{(*)} \tau \nu)}
	{\BF(B \to D^{(*)} \ell \nu)}\,,
	R_{K^{(*)}}\equiv \frac{\BF(B \to K^{(*)} \mu^+ \mu^-)}
	{\BF(B \to K^{(*)} e^+ e^-)}.
	\end{align}
	The precise measurement of those quantities would test the basic structure of the SM since LFU is only violated by the lepton masses in the SM.

	The world averaged experimental values~\cite{Amhis:2019ckw} based on measurements from 
	BaBar~\cite{Lees:2012xj}, Belle~\cite{Huschle:2015rga,Hirose:2016wfn,Belle:2019rba}, 
	and LHCb~\cite{Aaij:2015yra,Aaij:2017uff} are
	\begin{eqnarray}
	R_D=0.340 \pm 0.027 \pm 0.013\,~~~~R_{D^*}=0.295 \pm 0.011 \pm 0.008\,,
	\end{eqnarray}
	and the combined discrepancy to SM prediction is at the $3.1\sigma$ level
	~\cite{Altmannshofer:2020ywf,Amhis:2019ckw}.
	
	The most precise measurement to date of the $R_K$ has been performed by LHCb~\cite{Aaij:2021vac}
	\begin{eqnarray}
	R_K=0.846 ^{+0.044}_{-0.041}\,,~~~q^2\subseteq [1.1,6.0]~{\rm GeV^2}\,,
	\end{eqnarray}
	which has $3.1\sigma$ deviation from the SM expectation.
	For the $R_{K^*}$, LHCb Run-1 provides~\cite{Aaij:2017vbb}
	\begin{eqnarray}
	R_{K^*}=\left\lbrace
	\begin{array}{ll}
	0.66^{+0.11}_{-0.07} \pm 0.03\,,& q^2\subseteq [0.045,1.1]~{\rm GeV^2}\,,  \\ [2mm]
	0.69^{+0.11}_{-0.07} \pm 0.05\,,& q^2\subseteq [1.1,6.0]~{\rm GeV^2}\,.
	\end{array}
	\right.
	\end{eqnarray}
	Combining both $q^2$ bins, it has $2.5\sigma$ tension with the SM.
	On the other side, the $R_{K^*}$ and $R_K$ measurements from Belle
	~\cite{Abdesselam:2019wac,Abdesselam:2019lab}
	\begin{eqnarray}
	&& R_{K^*}=\left\lbrace
	\begin{array}{ll}
	0.90^{+0.27}_{-0.21} \pm 0.10\,,& q^2\subseteq [0.1,8.0]~{\rm GeV^2}\,,  \\ [2mm]
	1.18^{+0.52}_{-0.32} \pm 0.10\,,& q^2\subseteq [15,19]~{\rm GeV^2}\,,
	\end{array}
	\right. \nonumber \\ [2mm]
	&& R_{K}=\left\lbrace
	\begin{array}{ll}
	0.98^{+0.27}_{-0.23} \pm 0.06\,,& q^2\subseteq [1.0,6.0]~{\rm GeV^2}\,,  \\ [2mm]
	1.11^{+0.29}_{-0.26} \pm 0.07\,,& 14.18~{\rm GeV} < q^2\,,
	\end{array}
	\right.
	\end{eqnarray}
	are still compatible with SM predictions within their large uncertainties. 
	In the near future, Belle II is expected to significantly improve the uncertainties \cite{10.1093/ptep/ptz106}.

	\scp{The other {long-standing} problem is the anomalous magnetic moment of \yjk{muon}.} 
	Recently, the Muon $(g-2)$ experiment at Fermilab reported the value, 
	$a^{\rm FNAL}_\mu=(116592040 \pm 54)\times 10^{-11}$~\cite{PhysRevLett.126.141801}, or, \scp{the discrepancy from the SM} 
	\begin{eqnarray}
	\Delta a_\mu^{\rm FNAL} &=& a^{\rm FNAL}_\mu-a^{\rm SM}_\mu = 
	(230 \pm 69)\times 10^{-11}\,,
	\end{eqnarray} 
	which is a 3.3$\sigma$ \scp{deviation}. Since the value is compatible with the \yjk{earlier} value from BNL~\cite{Bennett:2006fi,Zyla:2020zbs}, the significance is \scp{now} strengthened to $4.2\sigma$ level:
	\begin{eqnarray}
	\Delta a_\mu^{\rm BNL+FNAL} &=& a^{\rm exp}_\mu-a^{\rm SM}_\mu = (251 \pm 59)\times 10^{-11}\,.
	\end{eqnarray}
	\scp{Even though it is not completely settled down from the lattice calculations~\cite{Borsanyi:2020mff}, certainly it is worth considering new physics as its solution. }

	Finding a common origin of the $B$-meson and $(g-2)_{\mu}$ anomalies is non-trivial,
	but it is appealing from the theoretical point of view.\footnote{See \cite{Jho:2019cxq, Jho:2020jsa, Ban:2020uii, Jho:2020sku, Park:2001xp, Park:2001uc, Kanemura:2015bli,Zhu:2021vlz,
Athron:2021iuf,Crivellin:2019dwb,Lindner:2016bgg,Terazawa:1968jh} for some of the earlier attempts to account $(g-2)$ and also various anomalies and possible experimental probes. } 
	In early attempts
~\cite{
Angelescu:2018tyl,Zyla:2020zbs,Buttazzo:2017ixm,
Aebischer:2019mlg,Cornella:2019hct,Assad:2017iib,
Alonso:2015sja,Calibbi:2015kma,Bhaskar:2021pml,
Dev:2020qet,Blanke:2018sro,Crivellin:2018yvo}, 
	the $U_1=(3,1)_{2/3}$ singlet vector Leptoquark, 
	in general couples to both left-handed (LH) and right-handed (RH) SM fermions,
	as single-mediator accounts for all the low-energy data.
	Its \yjk{simultaneous} explanations 
	of the $R_{K^{(*)}}$ and $R_{D^{(*)}}$ anomalies
	only \yjk{require} the LH couplings between second and third-generation quarks and leptons.
	However, the LH couplings cannot produce large enough 
	muon magnetic moment for $(g-2)_{\mu}$ anomaly.
	In this work, we further extend non-zero RH couplings of $U_1$, such that it substantially enhances the contribution to $(g-2)_{\mu}$.
	\scp{In addition, we will also consider $U(1)_{B_3-L_2}$ gauge boson ($X$) \cite{Bonilla:2017lsq, Alonso:2017uky, Allanach:2020kss} to  improve our fit to the experimental data.}
	We explored the plausible parameter space and search the common solution
	for both $B$-meson and $(g-2)_{\mu}$ anomalies.\footnote{
	 When we are finishing our paper, similar idea has been considered in \cite{Du:2021zkq}.  Unfortunately, however, they have missed some relevant constraints from low-energy experiments. In particular, the experimental $B_s \to \mu \mu$ data conflicts 
		with their preferred parameter region.
	}

	\bigskip
	
	\section{Model}
	
	\scp{In this section, we set our theoretical model to explain \yjk{$B$}-anomalies and $(g-2)_\mu$.} 
	
	\subsection{Vector Leptoquark $U_1=(3,1)_{2/3}$}
	\label{Sec2}
	\scp{Leptoquark is a natural candidate of new physics linking quark sector and lepton sector~\cite{Zyla:2020zbs}. 
	In particular,} we focus on the $U_1=(3,1)_{2/3}$ weak singlet vector leptoquark
	because it \scp{could provide simultaneous explanations 
	for $R_{K^{(*)}}$ and $R_{D^{(*)}}$ anomalies with its coupling with LH fermions
	~\cite{Angelescu:2018tyl,Buttazzo:2017ixm,Aebischer:2019mlg,Cornella:2019hct}.}
	However, the general Lagrangian includes the $U_1$ couplings to both LH and RH fermion under the SM gauge symmetry. 
	%However, to explain $R_{K^{(*)}}$ and $R_{D^{(*)}}$ anomalies,
	\scp{Including the most relevant interactions with the LH couplings to the 2nd and 3rd generations of leptons and quarks and also the RH couplings, we consider 
	the model Lagrangian:}
	\begin{eqnarray}
	\label{eq:L_U1}
	\mathcal{L} & \supset & U_{1\mu} \sum_{i,j=1,2,3}
	\Bigl [ x_L^{ij} (\bar{d}^i_L \gamma^\mu e^j_L )
	+\left( V_{\rm CKM} x_L U^{*}_{\rm PMNS} \right)_{ij} 
	(\bar{u}^i_L \gamma_\mu \nu^j_L) 
	+x^{ij}_R (\bar{d}^i_R \gamma^\mu e^j_R)
	\Bigr ]+h.c.
	\end{eqnarray}

	\scp{We adopt the real parts of CKM and PMNS matrices \yjk{which} is conveniently written as~\cite{Zyla:2020zbs}}
	\begin{eqnarray}
	V_{\rm CKM} x_L U^*_{\rm PMNS}=\left(
	\begin{array}{ccc}
	0.974~ & 0.225~ & 0.001 \\ [2mm]
	-0.224~ & 0.974~ & 0.042 \\ [2mm]
	0.009~ & -0.041~ & 0.999
	\end{array}
	\right)
	\left(
	\begin{array}{ccc}
	x^{11}_L & x^{12}_L & x^{13}_L \\ [2mm]
	x^{21}_L & x^{22}_L & x^{23}_L \\ [2mm]
	x^{31}_L & x^{32}_L & x^{33}_L
	\end{array}
	\right)
	\left(
	\begin{array}{ccc}
	0.821  & 0.551  & -0.150     \\ [2mm]
	-0.307 & 0.600  & 0.739     \\ [2mm]
	0.481  & -0.580 & 0.657
	\end{array}
	\right)\,, \nonumber 
	\end{eqnarray}
    above expression omits the imaginary parts in $V_{\rm CKM}$, but we adopt the full CKM parameterization from Ref~\cite{Workman:2022ynf}. in our numerical computation.
	The couplings $x^{ij}_L$ to the first generation leptons and quarks 
	are strongly constrained   from $\mu -e$ conversion on nuclei, and atomic parity violation 
	on $\BF(K \to \pi \nu \bar{\nu})$, \scp{ therefore we simply set them zero.}

	\scp{The most relevant Wilson coefficients of the effective Lagrangian in $R_{K^{(*)}}$, $R_{D^{(*)}}$, and $\BF(B_s \to \mu \mu)$ are  $C_9^{\mu\mu} (=-C_{10}^{\mu\mu})$~\cite{Angelescu:2018tyl}, and they are induced from the LH couplings}
	\begin{eqnarray}
	C^{\mu \mu}_9=-C^{\mu\mu}_{10}= -\frac{\pi v^2}{V_{tb} V^*_{ts}\alpha_{\rm EM}} \frac{x^{22}_L \left(x^{32}_L \right)^*}{m^2_{U_1}}\,,
	\end{eqnarray}
	where $v=246$ GeV is the vacuum expectation value of the Higgs, 
	and $\alpha_{\rm EM}$ is the fine structure constant.
	\scp{From the fit to $R_{K}$, $R_{K^*}$ and $\BF(B_s \to \mu \mu)$ data, we find  
	the parameter window for couplings and the mass of $U_1$}
	\begin{eqnarray}
	C^{\mu \mu}_9=-C^{\mu\mu}_{10}\subseteq [-0.85,-0.50]
	\Rightarrow -\frac{x^{22}_L (x^{32}_L)^*}{m^2_{U_1}}\subseteq [0.83,1.41]\times 10^{-3} ~{\rm TeV^{-2}}\,.
	\end{eqnarray}
	
	The interactions from Eq.(\ref{eq:L_U1}) also give rise to the effective coefficient~\cite{Angelescu:2018tyl}
	\begin{eqnarray}
	g_{V_L}=\frac{v^2}{2m^2_{U_1}}\left( x^{b\ell}_L \right)^*
	\left[ x^{b\ell'}_L+\frac{V_{cs}}{V_{cb}}x^{s\ell'}_L+\frac{V_{cd}}{V_{cb}}x^{d\ell'}_L \right]\,,
	\end{eqnarray}
	and contribute to $b \to c \ell \bar{\nu}_{\ell'}$.
	It becomes one solution for the $R_D$ and $R_{D^*}$ anomalies,
	and the $1\sigma$ region for $b \to c \tau {\overline{\nu}_\tau}$ requires
	\begin{eqnarray}
	g_{V_L}\subseteq [0.09,0.13]\Rightarrow \frac{(V^{\rm CKM}_{cs} x^{23}_L+V^{\rm CKM}_{cb}x^{33}_L)(x^{33}_L)^*}{m^2_{U_1}}\subseteq [0.12,0.18]~{\rm TeV^{-2}}\,.
	\end{eqnarray}

	\scp{The RH couplings $x_R$, combining with $x^{22}_L$ and $x^{32}_L$, contribute to the Wilson coefficients $(C_S)'=(C_P)'$, thus their magnitudes are bounded by the $B\to K\tau\mu$ and $B_s \to \mu\mu$ data, 
	which hampers from generating large enough muon magnetic dipole for $(g-2)_\mu$ anomaly.
%	thus
%	 are  constrained by the $b\to s X$ data, 
%	 which limits our explanation to the anomaly in $(g-2)_\mu$. 
	 Therefore, we are motivated to 
	 further extend our model.}

	\bigskip

    \subsection{Vector Leptoquark with  $U(1)_{B_3-L_2}$ $X$ boson}
    {
        In the \yjk{following} comprehensive analysis, with the preferred parameter region for $R_{K^{(*)}}$ and $R_{D^{(*)}}$ anomalies, the $U_1$ leptoquark itself may still not be  able to provide large enough contribution for $(g-2)_\mu$ anomaly.
        Therefore, we invoke the additional particle $X$, $U(1)_{B_3-L_2}$ gauge boson \scp{to enhance the contribution,}
        which is alternatively plausible candidate for $R_{K}$ anomaly \scp{and significantly change the preferred parameter space.}
    	We take the most relevant interactions with $X$ boson in the effective Lagrangian 
		\begin{align}
		\mathcal{L}_{\rm eff}^{X} & \supseteq  - g_X \bar{\mu} \gamma^\alpha \mu X_\alpha - g_X \bar{\nu}_\mu \gamma^\alpha P_L \nu_\mu X_\alpha  \nonumber \\
		&  + \frac{g_X}{3} \bar{{\bf u}}_L \gamma^\alpha \begin{pmatrix} 
		0 & 0 & 0 \\
		0 & 0 & 0 \\
		0 & 0 & 1
		\end{pmatrix} {\bf u}_L X_\alpha + \frac{g_X}{3} \bar{{\bf d}}_L \gamma^\alpha \underbrace{\begin{pmatrix} 
		| V_{td} |^2 & V_{ts} V_{td}^* & V_{tb} V_{td}^* \\
		V_{td} V_{ts}^* & | V_{ts} |^2 & V_{tb} V_{ts}^* \\
		V_{td} V_{tb}^* & V_{ts} V_{tb}^* & | V_{tb} |^2
		\end{pmatrix}}_{= V_{\rm CKM}^\dagger V_{B_3} V_{\rm CKM} } {\bf d}_L X_\alpha \ \ \ \ \ \ \\
		&+\frac{1}{2}m_X^2 X^\mu X_\mu \nonumber
		\end{align}
		where ${\bf u}^T_L\equiv (u_L,c_L,t_L)$ and ${\bf d}^T_L\equiv (d_L,s_L,b_L)$. { Here, we chose the configuration that 
the left-handed down-quark mixing contributes entire CKM matrix to avoid the constraint from $D-\bar{D}$ mixing.}
		\scp{We assume that the $m_X$ is induced by a new Higgs mechanism but the additional contribution from the new Higgs can be neglected.}
		The flavor-changing neutral current (FCNC) in down quarks from the $X$ boson 
		contributes to $B \to K \mu\mu$ and $B_s \to \mu \mu$. 
		At the same time, the muon coupling modifies the $(g-2)_\mu$. { In the following, we assume the $U(1)_{B_3-L_2}$ are broken under energy scale 
$\mathcal{O}(100)~{\rm GeV}$, 
which justifies the non-zero low-energy  
effective couplings at $\mathcal{O}(m_b)$ scale in Eq.~(\ref{eq:L_U1}).}
		%We will focus on $g_X\simeq \mathcal{O}(10^{-2})$ 
		%and $m_X=100~{\rm GeV}$, 
		%which makes $X$ boson contributions to $R_K$ substaintial, 
		%but to $(g-2)_\mu$ negligible.
    }

	\section{Low-energy observables}
	In this section, we summarize the low-energy observables with the $U_1$ vector leptoquark contributions.
	
	\subsection{$(g-2)_\mu$}
	The previous result of Muon $(g-2)$ experiment with the BNL E821 \yjk{was} 3.7$\sigma$ from the SM.  After the FNAL result, the \yjk{difference} between experiment and SM \yjk{has become}~\cite{Marciano:2016yhf,PhysRevLett.126.141801}
	\begin{eqnarray}
	\Delta a_\mu &=& a^{\rm exp}_\mu-a^{\rm SM}_\mu = (251\pm 59) \times 10^{-11}\,,
	\end{eqnarray}
\yjk{which is a deviation of} $4.2\sigma$ significance from the SM prediction and this \yjk{enhances} the motivation for SM extensions for new couplings with leptons.
	In this paper, we present the single leptoquark which is described in Sec.~\ref{Sec2} to explain not only $(g-2)_\mu$ anomaly but also $B$-meson anomalies.
	
	For the large mass of $U_1$ leptoquark, it contributes to the $(g-2)_\mu$ anomaly as\footnote{{
	We use the same $\kappa$ and $\tilde{\kappa}$ parameters defined in Ref.~\cite{Altmannshofer:2020ywf},
which are the tri-gauge boson couplings between $U_1$ leptoquarks and $B_\mu$ gauge boson from $U(1)_Y$. 
In Eq.~(\ref{eq:g_2}), we set $\kappa =1$ and $\tilde{\kappa}=0$, in such a way, 
the dipole moment becomes independent on the logarithmic term and UV theory.}}
	\begin{eqnarray}
	\Delta a_\mu = \frac{N_c}{16 \pi^2}&& \sum_i
	\left[ 2 Q_U \tilde{\kappa}_Y {\rm Im}(x_L^{i2}(x^{i2}_R)^*)\frac{m_{d_i} m_\mu}{m^2_{U_1}}\left(\text{ln}\left(\frac{\Lambda^2_{UV}}{M^2_{U1}}\right)+\frac{5}{2}\right)\right. \nonumber \\
	&&+2{\rm Re}(x^{i2}_L (x^{i2}_R)^*) \frac{m_{d_i} m_\mu}{m^2_{U_1}}
	\left(2Q_d+Q_{U_1} \left((1-\kappa_Y)\text{ln}\left(\frac{\Lambda^2_{UV}}{M^2_{U1}}\right)+\frac{1-5 \kappa_Y}{2} \right) \right)
	\nonumber \\
	&& \left. -(|x^{i2}_L|^2+|x^{i2}_R|^2)\frac{m^2_\mu}{m^2_{U_1}}
	\left( \frac{4}{3}Q_d+Q_{U_1}\left((1-\kappa_Y)\text{ln}\left(\frac{\Lambda^2_{UV}}{M^2_{U1}}\right) -\frac{1+9\kappa_Y}{6} \right) \right)  \right]\,.
	\end{eqnarray}
If $\kappa_Y\neq 1$ and $\tilde{\kappa}_Y\neq 0$, the dipole moment exhibits logarithmic dependence on the cut-off scale $\Lambda_{UV}$ not far above the leptoquark mass. So, the leptoquark contribution to $(g-2)_\mu$ anomaly becomes 
	\begin{eqnarray}
	\label{eq:g_2}
	\Delta a_\mu = \frac{N_c}{16 \pi^2} \sum_i
	&& \left[ 2{\rm Re}(x^{i2}_L (x^{i2}_R)^*) \frac{m_{d_i} m_\mu}{m^2_{U_1}}
	\left(2Q_d+Q_{U_1} \left(\frac{1-5 \kappa_Y}{2} \right) \right)
	\right. \nonumber \\
	&& \left. -(|x^{i2}_L|^2+|x^{i2}_R|^2)\frac{m^2_\mu}{m^2_{U_1}}
	\left( \frac{4}{3}Q_d+Q_{U_1}\left( -\frac{1+9\kappa_Y}{6} \right) \right)  \right]\,,
	\end{eqnarray}
	where we use $\kappa_Y=1$, $\tilde{\kappa}_Y=0$, $N_C = 3$, $Q_b=-1/3$, and $Q_{U_1}=2/3$. 
	We handle the renomalization group running from leptoquark scale down to muon mass by evaluating the quark masses at $\mathcal{O}$(TeV) scale in Eq.(\ref{eq:g_2}), 
	e.g. $m_b({\rm TeV})\simeq 2.4$ GeV 	~\cite{Altmannshofer:2020ywf}.

	\scp{
	The $X$ boson also \yjk{contributes} to $(g-2)_\mu$ as~\cite{Jho:2019cxq}
		\begin{align}
	\Delta a^X_\mu 
	&= \frac{g^2_X}{8 \pi^2} \int^1_0 dz 
	\frac{2z(1-z)^2}{(1-z)^2+z(m_X/m_\mu)^2} \\
	&\simeq (3g^2_X/4 \pi)(m^2_\mu/m^2_X)~~~\text{when $m_X\gg m_\mu$} \\
	&\simeq 2.7\times 10^{-11} \times \left(\frac{g_X}{0.01}\right)^2 \left(\frac{100~{\rm GeV}}{m_X}\right)^2,
	\end{align}
	thus it is negligible \yjk{compared} to the experimental value for $m_X\simeq 100$ GeV and $g_X\simeq 0.01$. 
	However, with $X$ boson contributions, the preferred parameter space for $B$-meson anomalies would move
	and our fit to the data can be significantly improved as we will describe below. 
	}
	
	\subsection{$R_{K^{(*)}}$, $R_{D^{(*)}}$}
	
	To explain the experimental result of $R_{K^{(*)}}$,
	it requires the Wilson coefficients as
	\begin{equation}
	\Delta C^{\mu\mu}_9|_{\rm exp}=-0.40\pm 0.12\,,\quad \Delta C^U_9|_{\rm exp}=-0.50\pm 0.38\,,
	\end{equation}
	with correlation $-0.5$~\cite{Aebischer:2019mlg,Alguero:2019ptt,Cornella:2019hct} between them.
	And the ratio of the SM predictions and experimental observation is 
	\begin{equation}
	\frac{R^{\rm exp}_D}{R^{\rm SM}_D}=1.14\pm 0.10 \,, \quad \frac{R^{\rm exp}_{D^*}}{R^{\rm SM}_{D^*}}=1.14\pm 0.05\,,
	\end{equation}
	with correlation $-0.37$ ~\cite{Cornella:2019hct}.
	For the $U_1$ leptoquark, the contribution to Wilson coefficients is
	~\cite{Aebischer:2019mlg,Cornella:2019hct}
	\begin{eqnarray}
	\label{eq:RK}
	&& \Delta C^{\mu\mu}_9=-\Delta C^{\mu\mu}_{10}=
	-\frac{4 \pi^2}{e^2} \frac{v^2}{m^2_{U_1}}\frac{x^{32}_L (x^{22}_L)^*}{V^*_{ts}V_{tb}}\,, \\
	\label{eq:CU9}
	&& \Delta C^{U}_9 \simeq -\frac{1}{V_{tb}V^*_{ts}}\frac{2}{3}
	\frac{v^2}{m^2_{U_1}} x^{23}_L (x^{33}_L)^* \ln\left( \frac{m^2_b}{m^2_{U_1}} \right)\,,
	\end{eqnarray}	
	where $\Delta C_9^U$ is the lepton-universal contribution 
    to $b \to s\ell \ell$ and originates from the $x^{23}_L$ 
    contributing through a log-enhanced photon penguin diagram~\cite{Cornella:2019hct,Crivellin:2018yvo}.
%	\sout{where $v\simeq 246$ GeV is the SM Higgs vacuum expectation value}. 
	Similarly, the $U_1$ contribution to $R_{D}$ and $R_{D^*}$ are~\cite{Cornella:2019hct}
	\begin{eqnarray}
	\label{eq:RD}
	\frac{R_D}{R^{\rm SM}_D}\simeq 
	\left[1+ \frac{v^2}{m^2_{U_1}}{\rm Re}
	\left\lbrace  \left(x^{33}_L-1.5\eta_S (x^{33}_R)^* \right)
	\frac{(V_{cb}x^{33}_L+V_{cs}x^{23}_L+V_{cd}x^{13}_L)}{V_{cb}} \right\rbrace 
	\right]\,, \nonumber \\
	\frac{R_{D^*}}{R^{\rm SM}_{D^*}}\simeq 
	\left[1+ \frac{v^2}{m^2_{U_1}}{\rm Re}
	\left\lbrace  \left(x^{33}_L-0.14\eta_S (x^{33}_R)^* \right)
	\frac{(V_{cb}x^{33}_L+V_{cs}x^{23}_L+V_{cd}x^{13}_L)}{V_{cb}} \right\rbrace 
	\right]\,,
	\end{eqnarray}
	where $\eta_S\simeq 1.8$ accounts for the running of the scalar operator from $m_{U1}=4$ TeV to $m_b$.

  	\scp{The additional correction from $X$ boson to the $\Delta C^{\mu\mu}_9$ 
	is given as~\cite{Allanach:2020kss}
    \begin{align}
    \label{eq:C9_X}
    \Delta C^X_9 
    &= \left( \frac{g^2_X}{3} V_{tb}V^*_{ts} \right)
    \left(\frac{36\,{\rm TeV}}{m_X} \right)^2  \nonumber \\
    &\approx -0.18 \times \left(\frac{g_X}{0.01}\right)^2 \left(\frac{100~{\rm GeV}}{m_X}\right)^2
    \end{align}
    It is important to notice that the negative value from $V_{ts}$ makes it \yjk{trend} toward 
    the experimental value for $B$-meson anomalies but does not contribute to $(g-2)_\mu$. }

	\subsection{$B^-_c \to \tau^- \overline{\nu}_\tau$, $B^+ \to \tau^+ \nu_\tau$} 
	
	%\yjk{\it The notation is very loose and needs improving: one is $\tau\bar{\nu}$ while the other is $\tau\nu$, and one is just $B_c$ while the other is $B^\pm$. My suggestion: $B_c^+ \to \tau^+ \nu_\tau,\ B^+ \to \tau^+ \nu_\tau$.  What do you think?}
	
	The bound from observation ~\cite{Akeroyd:2017mhr,Alonso:2016oyd}	and SM prediction for $B_c \to \tau \nu$ are~\cite{Altmannshofer:2020ywf,Blanke:2019qrx}
	\begin{eqnarray}
	\BF({B^-_c \to \tau^- \bar{\nu}_\tau})_{\rm exp}&&\leq 0.60 \nonumber \\
	\BF({B^-_c \to \tau^- \bar{\nu}_\tau})_{\rm SM}&&=(2.21 \pm 0.09)\times 10^{-2}
	\end{eqnarray}
	The proportion of experimental value and SM prediction for {$B^+ \to \tau^+ \nu_\tau$} is \cite{Altmannshofer:2020ywf}
	\begin{eqnarray}
	\frac{\BF({B^+ \to \tau^+ \nu_\tau})_{\rm exp}}{\BF({B^+ \to \tau^+ \nu_\tau})_{\rm SM}}=1.30 \pm 0.29\,.
	\end{eqnarray}	
	And the $U_1$ leptoquark contributions to each of obervables are \cite{Altmannshofer:2020ywf}
	\begin{equation}
	\label{eq:Bc_taunu}
	\frac{\BF(B^-_c \to \tau^- \bar{\nu}_\tau)}{\BF({B^-_c \to \tau^- \bar{\nu}_\tau})_{\rm SM}}=
	\left|1- \frac{(V_{cd}x^{13}_L+V_{cs}x^{23}_L+V_{cb}x^{33}_L)}{V_{cb}}
	\frac{v^2}{m^2_{U_1}} 
	\left( \frac{(x^{33}_L)^*}{2}+ \frac{(x^{33}_R)^* m^2_{B_c}}{m_\tau (m_b + m_c)} \right) \right|^2\,, 
	\end{equation}

	\begin{equation}
	\label{eq:B_taunu}
	\frac{\BF(B^+ \to \tau^+ \nu_\tau)}{\BF({B^+ \to \tau^+ \nu_\tau})_{\rm SM}}=
	\left|1- \frac{(V_{ud}x^{13}_L+V_{us}x^{23}_L+V_{ub}x^{33}_L)}{V_{ub}}
	\frac{v^2}{m^2_{U_1}} 
	\left( \frac{(x^{33}_L)^*}{2}+ \frac{(x^{33}_R)^* m^2_{B^\pm}}{m_\tau m_b} \right) \right|^2\,.
	\end{equation}
	
	\subsection{{$B^0_s \to \tau^+ \tau^-$, $B^0_s \to \mu^+ \mu^-$, $B^0_s \to \tau^\pm \mu^\mp$ and $B^+ \to K^+ \tau^+ \tau^-$}}

	The experimental value from LHCb is~\cite{Aaij:2017xqt} and SM prediction is~\cite{Bobeth:2013uxa}
	\begin{eqnarray}
	\BF(B^0_s \to \tau^+ \tau^-)_{\rm exp}&{<6.8 \times 10^{-3}~~\text{at 95\% C.L}}\,,\\
	\BF(B^0_s \to \tau^+ \tau^-)_{\rm SM}&=(7.73\pm 0.49)\times 10^{-7}\,.
	\end{eqnarray}
	And the related contribution for $U_1$ leptoquark is \cite{Altmannshofer:2020ywf}
	\begin{eqnarray}
	\label{eq:Bs_tautau}
	\frac{\BF(B^0_s \to \tau^+ \tau^-)}{\BF({B^0_s} \to \tau^+ \tau^-)_{\rm SM}}  
	&=&
	\frac{16 \pi^4}{e^4 (C^{\rm SM}_{10})^2}\frac{v^4}{m^4_{U_1}}
	\frac{m^4_{B_s}}{m^2_\tau m^2_b}
	\left| \frac{(x^{33}_L)^*x^{23}_R-(x^{33}_R)^*x^{23}_L}{V^*_{ts}V_{tb}} \right|^2
	\left(1-\frac{4m^2_\tau}{m^2_{B_s}} \right)
	\nonumber \\
	&& + \left|1+\frac{4\pi^2}{e^2 C^{\rm SM}_{10}} \frac{v^2}{m^2_{U_1}}
	\left( \frac{(x^{33}_L)^* x^{23}_L+(x^{33}_R)^*x^{23}_R}{V^*_{ts}V_{tb}} 
	- \frac{m^2_{B_s}}{m_\tau m_b}
	\frac{(x^{33}_L)^*x^{23}_R+(x^{33}_R)^*x^{23}_L}{V^*_{ts}V_{tb}} \right)
	\right|^2\,, \nonumber \\
	\end{eqnarray}
	where $C^{\rm SM}_{10}\simeq -4.1$ which we use for a normalization such that the SM value for the Wilson coefficient ~\cite{Altmannshofer:2008dz}.

	The ratio between the SM prediction and experimental value is~\cite{Altmannshofer:2020ywf,Aebischer:2019mlg}
	\begin{align}
	\frac{\BF({B^0_s} \to \mu^+ \mu^-)_{\rm exp}}{\BF({B^0_s} \to \mu^+ \mu^-)_{\rm SM}} =0.73^{+0.13}_{-0.10}\,.
	\end{align}	
	And the following $U_1$ vector leptoquark and $X$-boson total contribution is written by~\cite{Altmannshofer:2020ywf,Cheung:2006tm}
	\begin{align}
	\label{eq:Bs_mumu}
	\frac{\BF({B^0_s} \to \mu^+ \mu^-)}{\BF({B^0_s} \to \mu^+ \mu^-)_{\rm SM}}  
	&=
	\frac{16 \pi^4}{e^4 (C^{\rm SM}_{10})^2}\frac{v^4}{m^4_{U_1}}
	\frac{m^4_{B_s}}{m^2_\mu m^2_b}
	\left| \frac{(x^{32}_L)^*x^{22}_R-(x^{32}_R)^*x^{22}_L}{V^*_{ts}V_{tb}} \right|^2
	\nonumber \\
	& + \left|1+\frac{4\pi^2}{e^2 C^{\rm SM}_{10}} \frac{v^2}{m^2_{U_1}}
	\left( \frac{(x^{32}_L)^* x^{22}_L+(x^{32}_R)^*x^{22}_R}{V^*_{ts}V_{tb}} 
	- \frac{m^2_{B_s}}{m_\mu m_b}
	\frac{(x^{32}_L)^*x^{22}_R+(x^{32}_R)^*x^{22}_L}{V^*_{ts}V_{tb}} \right) \right. \nonumber \\
	&~~~~~~ \left. -\left[ \frac{\alpha}{2 \pi \sin^2 \theta_W}Y\left( \frac{m^2_t}{m^2_W} \right) \right]^{-1} 
\left(
\frac{2\,g^2_X\,m^2_Z}{3\, g^2\, m^2_X} 
\right)
	\right|^2 \,,
	\end{align}
	where the last term in the second absolute bracket comes from the $X$-boson.
	Here we take $Y(m^2_t/m^2_W)=1.05$, and $g\simeq 0.652$ the $SU(2)_L$ coupling constant.
	
	%%%KY

		LHCb search on $B_s^0 \to \tau^\pm \mu^\mp$ provides an upper limit
		\begin{eqnarray}
		\BF(B_s^0 \to \tau^\pm \mu^\mp)_{\rm exp} < 2.1 \times 10^{-5},
		\end{eqnarray}
		at 95 \% confidence level \cite{Aaij:2019okb}. SM prediction of this branching fraction is extremely small as $\mathcal{O}(10^{-54})$ \cite{Calibbi:2017uvl}. The expression of $U_1$ contribution to \yjk{$B_s^0 \to \tau^\pm \mu^\mp$} is~\cite{Cornella:2019hct}
		\begin{eqnarray}
		\label{eq:Bs_taumu}
		\BF({B^0_s \to \tau^\pm \mu^\mp})= 
		\frac{1}{\Gamma_{B_s}}&&\frac{m_{B_s}f^2_{B_s}G^2_F}{8\pi}
		m^2_\tau \left(1-\frac{m^2_\tau}{m^2_{B_s}} \right)^2\nonumber \\
		&&\times \frac{v^4}{4m^4_{U_1}}
		\left|x^{22}_L(x^{33}_L)^*-\frac{2\eta_S m^2_{B_s}}{m_\tau(m_s+m_b)}
		x^{22}_L(x^{33}_R)^* \right|^2\,. 
		\end{eqnarray} 
		where $G_F=1.166\times 10^{-5}~{\rm GeV^{-2}}$ is the Fermi constant, $f_{B_s}=0.225$ GeV~\cite{McNeile:2011ng} is the leptonic decay constant of $B^0_s$, and 
		$\Gamma_{B_s}=4.34\times 10^{-13}~{\rm GeV}$ is the total width of $B^0_s$.

		BaBar experiment measured the branching fraction \cite{TheBaBar:2016xwe}
		\begin{eqnarray}
		\BF(B^+ \to K^+ \tau^+ \tau^-)_{\rm exp}=(1.31\pm 0.71)\times 10^{-3}\,.
		\end{eqnarray}
		with an upper limit of $\text{Br}(B^+ \to K^+ \tau^+ \tau^-) < 2.25 \times 10^{-3}$ at the $90\%$ confidence level. The expression of $U_1$ leptoquark contribution to this process is given by~\cite{Cornella:2019hct}
		\begin{eqnarray}
		\label{eq:B_ktautau}
		\BF(B^+ \to K^+ \tau^+ \tau^-)\simeq &&
		1.5\times 10^{-7}+10^{-3} \frac{v^2}{2m^2_{U_1}}
		\left[1.4\,{\rm Re}\left( x^{23}_L (x^{33}_L)^* \right)
		-3.3\,{\rm Re}\left(x^{23}_L (x^{33}_R)^* \right) \right] \nonumber \\
		&&+\frac{v^4}{4m^4_{U_1}}\left| x^{23}_L \right|^2
		\left[ 3.5 \left| x^{33}_L\right|^2
		-16.4\,{\rm Re}\left( x^{33}_R (x^{33}_L)^* \right)
		+95.0 \left| x^{33}_R \right|^2 \right]\,.
		\end{eqnarray}
		where $v=246$ GeV is the electroweak vacuum expectation value.

	\subsection{$B^+ \to K^+ \tau^+ \mu^-$, $B^+ \to K^+ \tau^- \mu^+$}

		From BaBar experiment, we obtain upper limits~\cite{Lees:2012zz}
		\begin{eqnarray}
		\BF(B^+ \to K^+ \tau^+ \mu^-) & < & 2.8\times 10^{-5}\,, \\
		\BF(B^+ \to K^+ \tau^- \mu^+) & < & 4.5\times 10^{-5}\,
		\end{eqnarray}
		at 90\% confidence level. The leptoquark contribution is given by~\cite{Cornella:2019hct,Becirevic:2016zri}
		\begin{eqnarray}
		\label{eq:B_ktaumu}
		&& \BF(B^+ \to K^+ \tau^+ \mu^-)\simeq
		\frac{v^4}{4m^4_{U_1}}\left| x^{22}_L \right|^2
		\left[8.3\left| x^{33}_L \right|^2 + 155.2\left| x^{33}_R \right|^2-42.3\,{\rm Re}\left((x^{33}_L)^* x^{33}_R\right)  \right],
		\nonumber \\
		&&\BF(B^+ \to K^+ \tau^- \mu^+)\simeq 8.3 \frac{v^4}{4m^4_{U_1}}\left|x^{32}_L (x^{23}_L)^* \right|^2\,.
		\end{eqnarray}

	\subsection{$\tau \to \mu \gamma,\ \mu \phi$ and LFU in $\tau$ \yjk{decays}}

		Due to its sizeable couplings to muon and tau leptons, $U_1$ leptoquark can significantly affect the Lepton-flavor-violation in $\tau$ decays. The experimental upper limits are \cite{Amhis:2016xyh, Miyazaki:2011xe}
		\begin{eqnarray}
		\BF(\tau \to \mu \gamma) & < & 3.0 \times 10^{-8}\,. \\
		\BF(\tau \to \mu \phi) & < & 5.1 \times 10^{-8}\,
		\end{eqnarray}
		at 90\% confidence level. In addition, the LFU in the decay of charged leptons can give stringent bounds on the leptoquark couplings. The experimentally measured values are \cite{Amhis:2016xyh,Bordone:2018nbg}
		\begin{eqnarray}
		&&(g_\tau/g_\mu)_{\rm exp}= 1.0000 \pm 0.0014, \\
		&&(g_\tau/g_\mu)_{\ell}= 1.0010 \pm 0.0015, \\
		&&(g_\tau/g_\mu)_{\pi}= 0.9961 \pm 0.0027, \\
		&&(g_\tau/g_\mu)_{K}= 0.9860 \pm 0.0070\,.
		\end{eqnarray}
		
		For the $U_1$ leptoquark contribution to $\BF(\tau \to \mu \gamma)$, %\cite{Altmannshofer:2020ywf}
%		\begin{eqnarray}
%		\label{eq:tau_mug}
%		\BF(\tau \to \mu \gamma)\simeq \frac{1}{\Gamma_\tau}
%		\frac{\alpha_{\rm EM}N^2_c}{256 \pi^4} 
%		\frac{m^3_\tau m^2_b}{m^4_{U_1}} \left|x^{33}_R (x^{32}_L)^* \right|^2
%		\left[2Q_b-Q_{U_1}\left( \frac{1-5 \kappa_Y}{2} \right)  \right]^2\,,
%		\end{eqnarray}
%		where we use $\kappa_Y=1$, $\tilde{\kappa}_Y=0$, $Q_b=-1/3$, and %$Q_{U_1}=2/3$, $\Gamma_\tau=2.27\times 10^{-12}~{\rm GeV}$. 
we adopt complete formula of the decay width~\cite{Hati:2019ufv}
\begin{eqnarray}
\label{eq:taumua_complete1}
\Gamma(\tau \to \mu \gamma)=\frac{\alpha_{\rm EM}(m^2_\tau-m^2_\mu)^3}{4m^3_\tau}
\left( \left| \sigma^{32}_L \right|^2 + \left| \sigma^{32}_R \right|^2  \right)\,,
\end{eqnarray}		
where
\begin{eqnarray}
\label{eq:taumua_complete2}
\sigma^{32}_L = -\frac{iN_c}{16\pi^2 m^2_{U_1}}
\sum_{k=1,2,3}
&& \left\lbrace 
\frac{2}{3}\left[
\left( {x^{k2}_R}^*x^{k3}_R m_\tau + {x^{k2}_L}^*x^{k3}_L m_\mu  \right) g(t_k)
+{x^{k2}_R}^*x^{k3}_L m_{d_k}j(t_k)
\right] \right. \nonumber \\
&& \left. -\frac{1}{3}\left[
\left( {x^{k2}_R}^*x^{k3}_R m_\tau + {x^{k2}_L}^*x^{k3}_L m_\mu  \right) f(t_k)
+{x^{k2}_R}^*x^{k3}_L m_{d_k}h(t_k)
\right]
\right\rbrace \nonumber \\
\sigma^{32}_R = -\frac{iN_c}{16\pi^2 m^2_{U_1}}
\sum_{k=1,2,3}
&& \left\lbrace 
\frac{2}{3}\left[
\left( {x^{k2}_L}^*x^{k3}_L m_\tau + {x^{k2}_R}^*x^{k3}_R m_\mu  \right) g(t_k)
+{x^{k2}_L}^*x^{k3}_R m_{d_k}j(t_k)
\right] \right. \nonumber \\
&& \left. -\frac{1}{3}\left[
\left( {x^{k2}_L}^*x^{k3}_L m_\tau + {x^{k2}_R}^*x^{k3}_R m_\mu  \right) f(t_k)
+{x^{k2}_L}^*x^{k3}_R m_{d_k}h(t_k)
\right]
\right\rbrace \nonumber \\
\end{eqnarray}
with $t_k\equiv m^2_{d_k}/m^2_{U_1}$ and $f,g,h,j$ are loop functions~\cite{Hati:2019ufv}.

		For $\BF(\tau \to \mu \phi)$, it is given by \cite{Bhattacharya:2016mcc,Cornella:2019hct}
		\begin{eqnarray}
		\label{eq:tau_muphi}
		\BF(\tau \to \mu \phi)=
		\frac{1}{\Gamma_\tau} \frac{f^2_\phi G^2_F}{16 \pi}m^3_\tau 
		\left(1-\frac{m^2_\phi}{m^2_\tau} \right)^2
		\left(1+\frac{2m^2_\phi}{m^2_\tau} \right)
		\frac{v^4}{4m^4_{U_1}}\left| x^{23}_L (x^{22}_L)^* \right|^2\,,
		\end{eqnarray}
		where $\phi$ is the $s\bar{s}$ vector meson with $f_\phi=0.225$ GeV, $m_\phi=1.019~{\rm GeV}$~\cite{Bhattacharya:2016mcc}. For LFU in lepton decays, we use the expression \cite{Bordone:2018nbg,Cornella:2019hct}
		\begin{eqnarray}
		\left( \frac{g_\tau}{g_\mu} \right)_{\ell,\pi,K}\simeq 
		1-0.08 \times \frac{(x^{33}_L)^2 v^2}{4m^2_{U1}}\,.
		\end{eqnarray}
		
		More specifically, they can be written in terms of the effective Lagrangian for leptonic decay~\cite{Bordone:2018nbg}
		\begin{eqnarray}
		\mathcal{L}_{\ell \to \ell' \nu \bar{\nu}}
		=-\frac{4G_F}{\sqrt{2}}
		\left( 
		\left[C^{\rm V,LL}_{\nu e} \right]_{\rho \sigma \alpha \beta}
		(\bar{\nu}^\rho_L \gamma^\mu \nu^\sigma_L )
		(\bar{\ell}^\alpha_L \gamma^\mu \ell^\beta_L )
		+\left[C^{\rm V,LR}_{\nu e} \right]_{\rho \sigma \alpha \beta}
		(\bar{\nu}^\rho_L \gamma^\mu \nu^\sigma_L )
		(\bar{\ell}^\alpha_R \gamma^\mu \ell^\beta_R )
		\right)\,, \nonumber \\
		\end{eqnarray}
		and for hadronic decay~\cite{Bordone:2018nbg}
		\begin{eqnarray}
		\mathcal{L}_{\tau \to h \nu}
		=-\frac{4G_F}{\sqrt{2}} \sum_\rho
		\left\lbrace
		\left( 
		\delta_{\rho 3} (V_{\rm CKM})^*_{ji}
		\left[C^{\rm V,LL}_{\nu e du} \right]_{\rho 3 i j}
		\right)
		(\bar{\nu}^\rho_L \gamma^\mu \tau_L )
		(\bar{d}^i_L \gamma^\mu u^j_L )
		+\left[C^{\rm S,RL}_{\nu e du} \right]_{\rho 3 ij}
		(\bar{\nu}^\rho_L  \tau_L )
		(\bar{d}^i_R  u^j_R ) 
		\right\rbrace\,
		\nonumber \\
		\end{eqnarray}
		result in the following expressions for couplings with $\tau$ and $\mu$
		\begin{eqnarray}
		\label{eq:LFU}
		&& \left( \frac{g_\tau}{g_\mu} \right)_\ell
		=
		\left[
		\frac{\sum_{\rho \sigma} 
			\left(
			\left|\delta_{\rho 3}\delta_{\sigma 1}+ \left[ C^{\rm V,LL}_{\nu e} \right]_{\rho \sigma 13} \right|^2+\left| \left[ C^{\rm V,LR}_{\nu e} \right]_{\rho \sigma 13} \right|^2 
			\right) }
		{\sum_{\rho \sigma} 
			\left(
			\left|\delta_{\rho 2}\delta_{\sigma 1}+ \left[ C^{\rm V,LL}_{\nu e} \right]_{\rho \sigma 12} \right|^2+\left| \left[ C^{\rm V,LR}_{\nu e} \right]_{\rho \sigma 12} \right|^2 
			\right) }
		\right]^{1/2} \,, \nonumber \\
		&& \left( \frac{g_\tau}{g_\mu} \right)_\pi
		=
		\left(
		\frac{\sum_{\rho} 
			\left|\delta_{\rho 3}V^*_{ud}+ \left[ C^{\rm V,LL}_{\nu e du} \right]_{\rho 311} 
			+\frac{m^2_\pi}{m_\tau(m_d+m_u)} \left[ C^{\rm S,RL}_{\nu e du} \right]_{\rho 311} \right|^2 
		}
		{\sum_{\rho} 
			\left|\delta_{\rho 2}V_{ud}+ \left[ C^{\rm V,LL}_{\nu e du} \right]^*_{\rho 211} 
			+\frac{m^2_\pi}{m_\mu(m_d+m_u)} \left[ C^{\rm S,RL}_{\nu e du} \right]^*_{\rho 211} \right|^2 
		}
		\right)^{1/2} \,, \nonumber \\
		&& \left( \frac{g_\tau}{g_\mu} \right)_K
		=
		\left(
		\frac{\sum_{\rho} 
			\left|\delta_{\rho 3}V^*_{us}+ \left[ C^{\rm V,LL}_{\nu e du} \right]_{\rho 321} 
			+\frac{m^2_K}{m_\tau(m_d+m_u)} \left[ C^{\rm S,RL}_{\nu e du} \right]_{\rho 321} \right|^2 
		}
		{\sum_{\rho} 
			\left|\delta_{\rho 2}V_{us}+ \left[ C^{\rm V,LL}_{\nu e du} \right]^*_{\rho 221} 
			+\frac{m^2_K}{m_\mu(m_d+m_u)} \left[ C^{\rm S,RL}_{\nu e du} \right]^*_{\rho 221} \right|^2 
		}
		\right)^{1/2} \,.
		\end{eqnarray}

		For $U_1$ leptoquark, by using the Fierz transformation in Eq.(\ref{eq:L_U1}), 
		$$
		[\bar{u}_{1L} \gamma^\mu u_{2L}][\bar{u}_{3L} \gamma_\mu u_{4L}]=
		-[\bar{u}_{1L} \gamma^\mu u_{4L}][\bar{u}_{3L} \gamma_\mu u_{2L}]\,,
		$$
		we replace the Wilson coefficients as
		\begin{eqnarray}
		&& \left[ C^{\rm V,LL}_{\nu e du} \right]_{\rho \sigma i j}
		\Leftrightarrow
		-\frac{2v^2}{4 m^2_{U_1}}
		\left(V_{\rm CKM}x_L U^*_{\rm PMNS}\right)_{j \rho}(x_L)_{i\sigma}\,, \nonumber \\
		&&\left[ C^{\rm S,RL}_{\nu e du} \right]=
		\left[ C^{\rm V,LL}_{\nu e} \right]
		=\left[ C^{\rm V,LR}_{\nu e} \right]
		=0\,.
		\end{eqnarray}

	\begin{table}[]
		\centering
		\caption{\label{tab1} The list of observables for the $\chi^2$ scanning with the measured values and predictions from SM. The equations of the $U_1(+X)$ model are also referenced. }
		\begin{tabular}{c||c|c|c}
			\hline\hline
			Observable & Experiment & SM predict & $U_1(+X)$ predict \\ \hline\hline
			\addstackgap[15pt]{$R_{D^{(*)}}$} & \multicolumn{2}{c|}{$\cfrac{R^{\text{exp}}_D}{R^{\rm SM}_D} = 1.14\pm 0.10,\quad$  $\cfrac{R^{\text{exp}}_{D^*}}{R^{\rm SM}_{D^*}} = 1.14\pm 0.05$} & (\ref{eq:RD}) \\ \hline\hline
			$\Delta \mathcal{C}_9^{\mu\mu}=-\Delta \mathcal{C}_{10}^{\mu\mu}$ ($R_{K^{(*)}}$) & $-0.40\pm 0.12$ & 0 & (\ref{eq:RK}) \\ \hline
			$\Delta \mathcal{C}_9^{U}$ & $-0.50 \pm 0.38$ & 0 & (\ref{eq:CU9}), (\ref{eq:C9_X}) \\ \hline\hline
			$B^-_c\rightarrow \tau^- \bar{\nu}_\tau$ & $\leq$ 0.60 & (2.21$\pm$0.09)$\times 10^{-2}$ & (\ref{eq:Bc_taunu}) \\ \hline
			$B^+\rightarrow \tau^+\nu_\tau$ & (1.09$\pm$0.24)$\times 10^{-4}$ & (8.8$\pm$0.6)$\times 10^{-5}$ & (\ref{eq:B_taunu}) \\ \hline
			$B^0_s\rightarrow \tau^+\tau^-$ & {$< 6.8\times 10^{-3}$} & $(7.73\pm 0.49)\times 10^{-7}$ & (\ref{eq:Bs_tautau}) \\ \hline
			\addstackgap[15pt]{$B^0_s\rightarrow \mu^+\mu^-$} & \multicolumn{2}{c|}{$\cfrac{BR(B^0_s\rightarrow \mu^+\mu^-)^{\text{exp}}}{BR(B^0_s\rightarrow \mu^+\mu^-)^{\rm SM}} = 0.73^{+0.13}_{-0.10}$} & (\ref{eq:Bs_mumu}) \\ \hline
			$B^0_s\rightarrow \tau^\pm \mu^\mp$ & {$< 2.1\times 10^{-5}$} &  & (\ref{eq:Bs_taumu}) \\ \hline
			$B^+\rightarrow K^+\tau^+\tau^-$ & $(1.31\pm 0.71)\times 10^{-3}$ & $(1.20\pm 0.12)\times 10^{-7}$ & (\ref{eq:B_ktaumu}) \\ \hline
			$B^+\rightarrow K^+\tau^+\mu^-$ & $\leq 2.8\times 10^{-5}$ &  & (\ref{eq:B_ktaumu}) \\ \hline
			$B^+\rightarrow K^+\tau^-\mu^+$ & $\leq 4.5\times 10^{-5}$ &  &  (\ref{eq:B_ktaumu}) \\ \hline\hline
			$\tau \rightarrow \mu\gamma$ & {$< 3.0\times 10^{-8}$} &  & (\ref{eq:taumua_complete1}) \\ \hline
			$\tau \rightarrow \mu\phi$ & {$< 5.1\times 10^{-8}$} &  & (\ref{eq:tau_muphi})  \\ \hline
			\multirow{3}{*}{LFU in $\tau$ decay} & $(g_\tau/g_\mu)_{\ell}= 1.0010 \pm 0.0015$ &  & \multirow{3}{*}{(\ref{eq:LFU})} \\ \cline{2-3}
			& $(g_\tau/g_\mu)_{\pi}= 0.9961 \pm 0.0027$ &  &  \\ \cline{2-3}
			& $(g_\tau/g_\mu)_{K}= 0.9860 \pm 0.0070$ &  &  \\ \hline\hline
		\end{tabular}
	\end{table}

\newpage

	\section{Parameter scanning}

\begin{figure}[h]
		\centering
		\includegraphics[height=1.5in,angle=0]{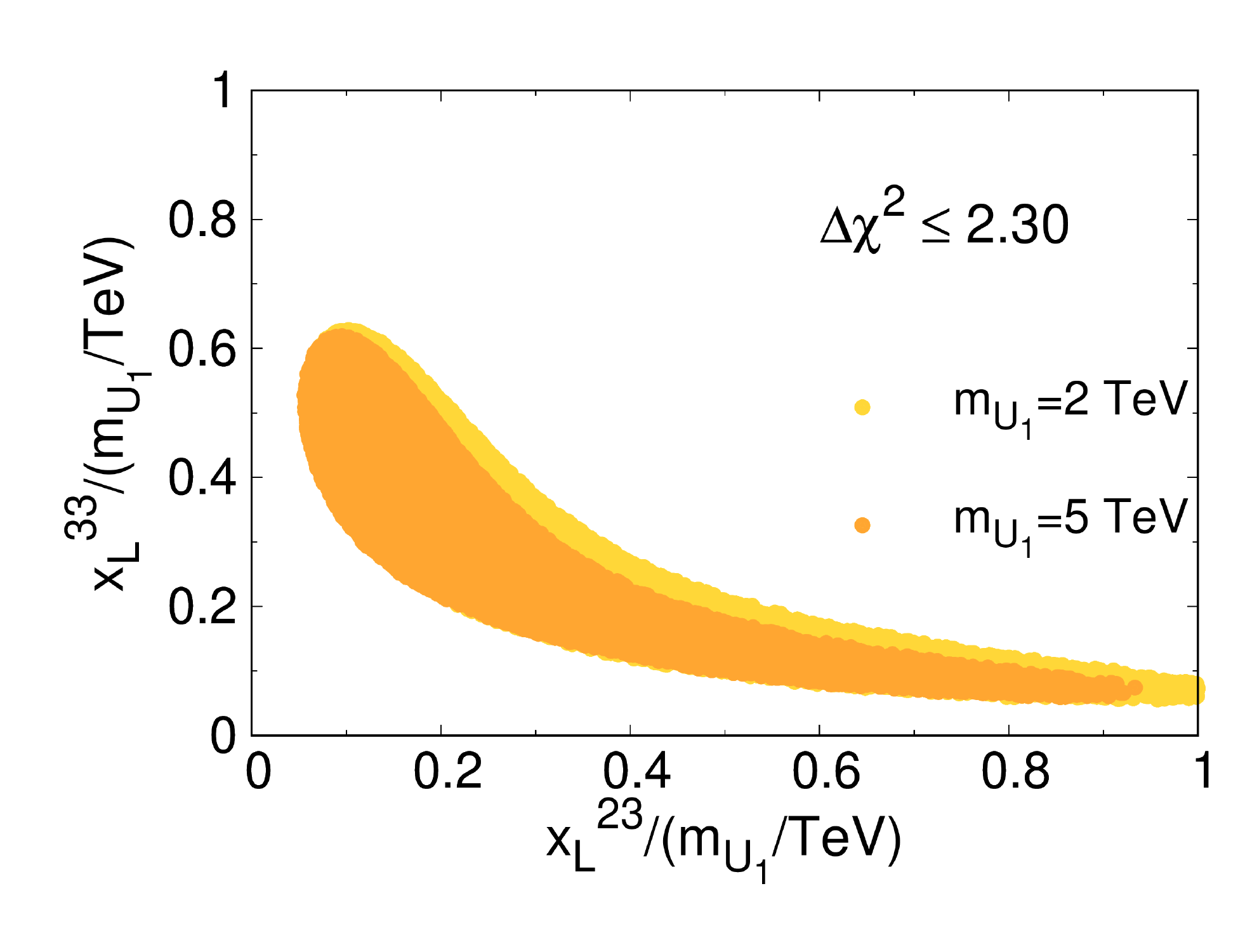}
		\includegraphics[height=1.5in,angle=0]{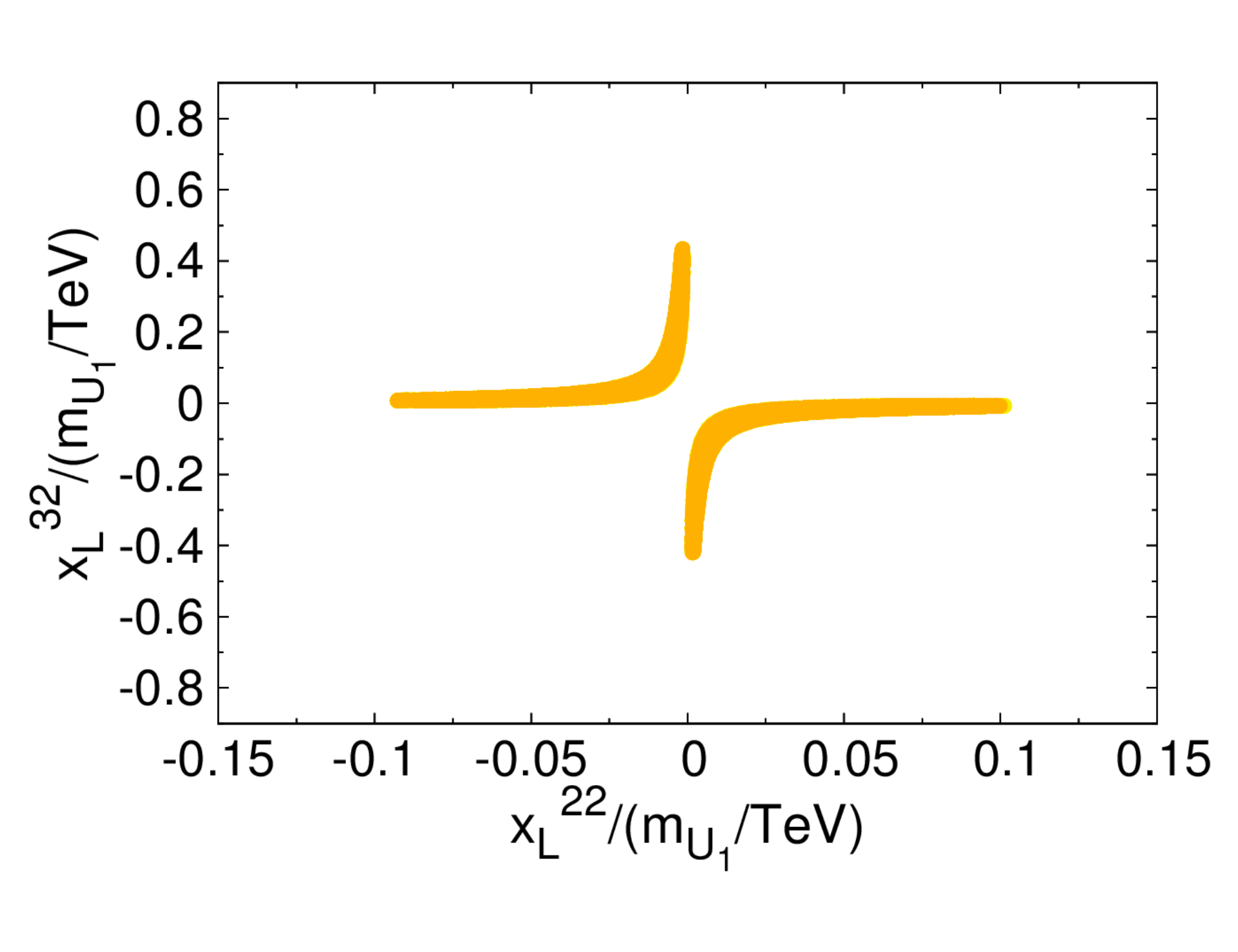}
		\includegraphics[height=1.5in,angle=0]{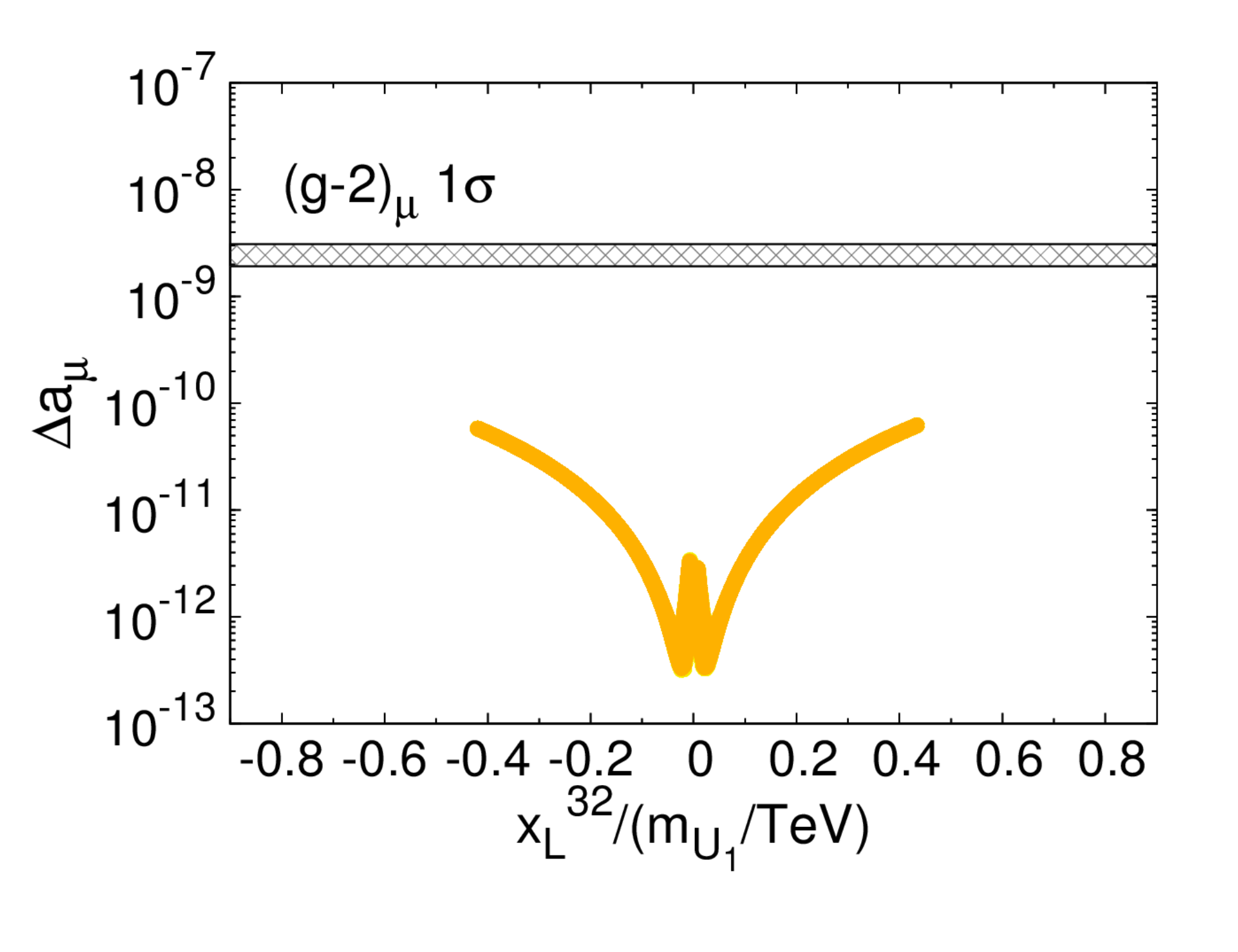}
		\includegraphics[height=1.5in,angle=0]{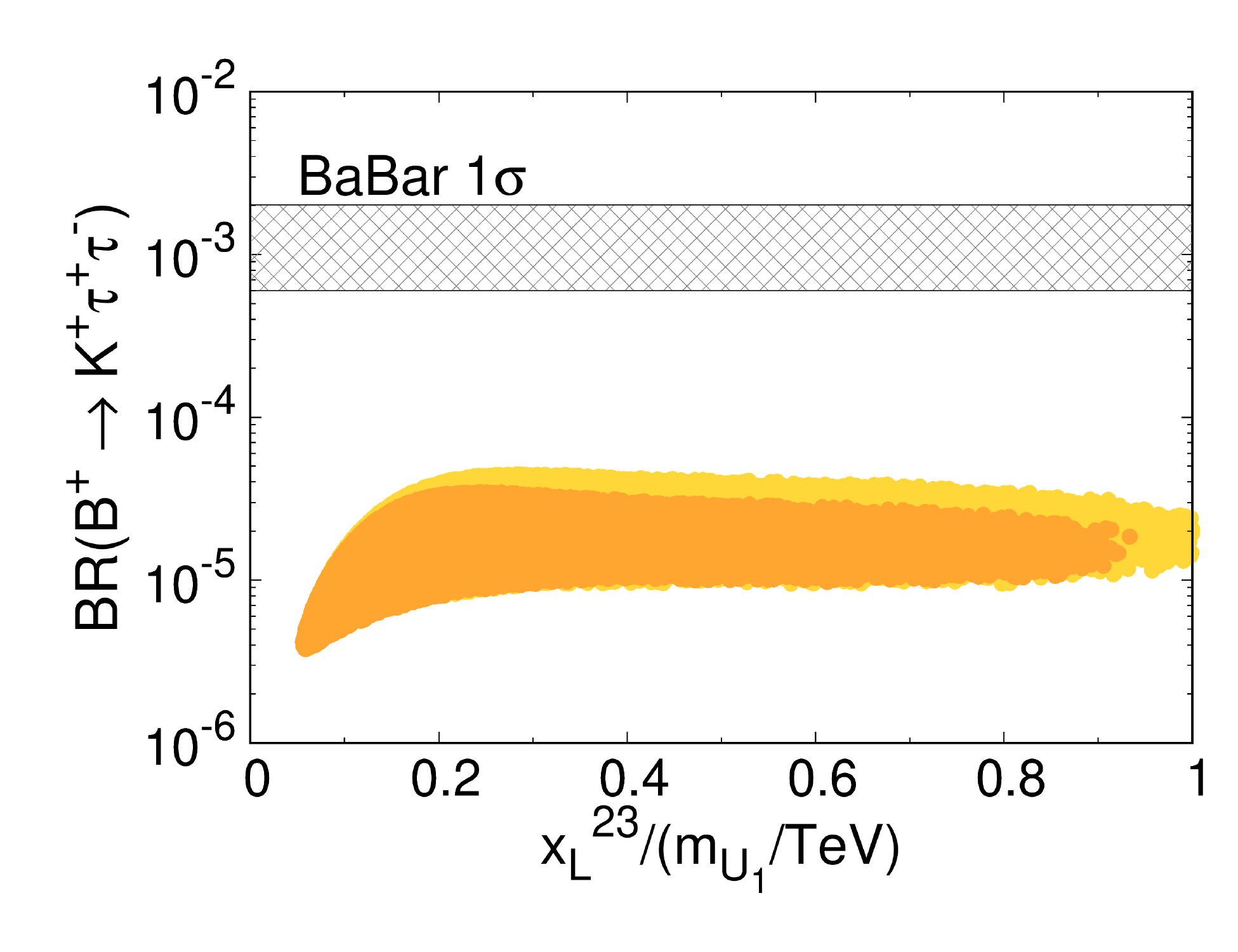}
		\includegraphics[height=1.5in,angle=0]{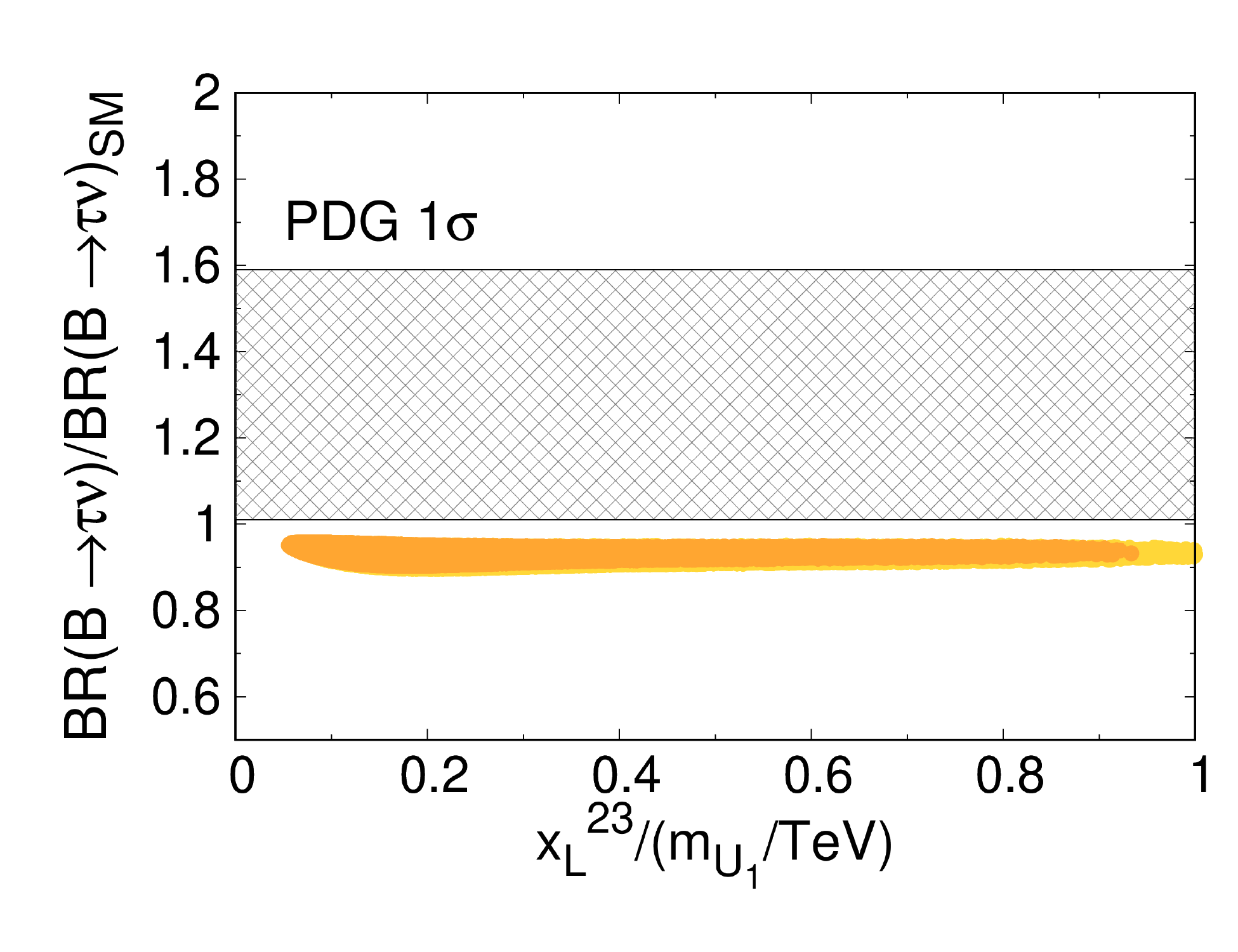}
		\includegraphics[height=1.5in,angle=0]{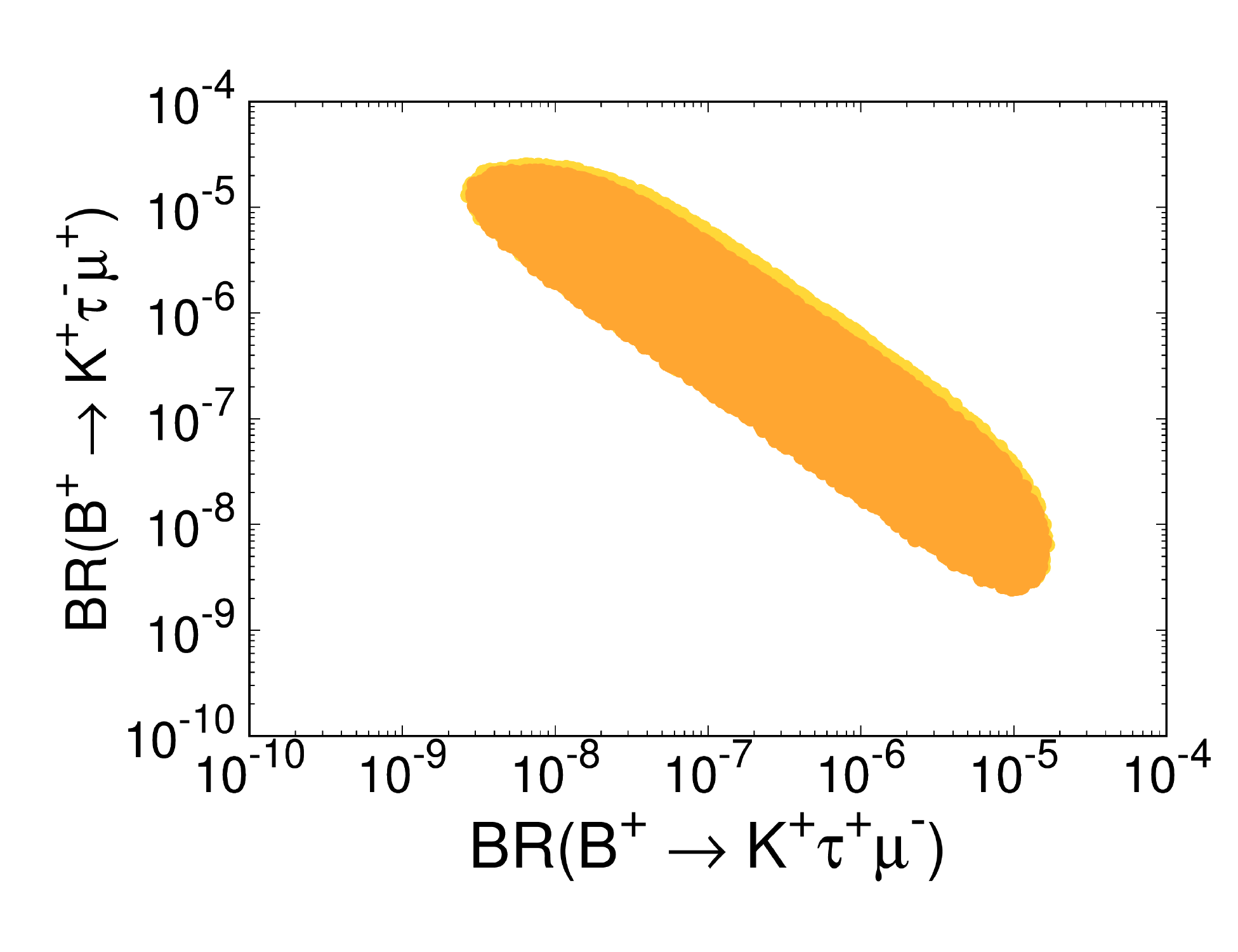}
		\caption{\small \label{fig:scan1}
			{\bf Scan-1:} The region satisfies $\Delta \chi^2 \equiv \chi^2 - \chi^2_{\rm min,1}\leq 2.3$, and $\chi^2_{\rm min,1}=9.23$. }
	\end{figure}

\begin{figure}[ht]
		\centering
	         \includegraphics[height=1.5in,angle=0]{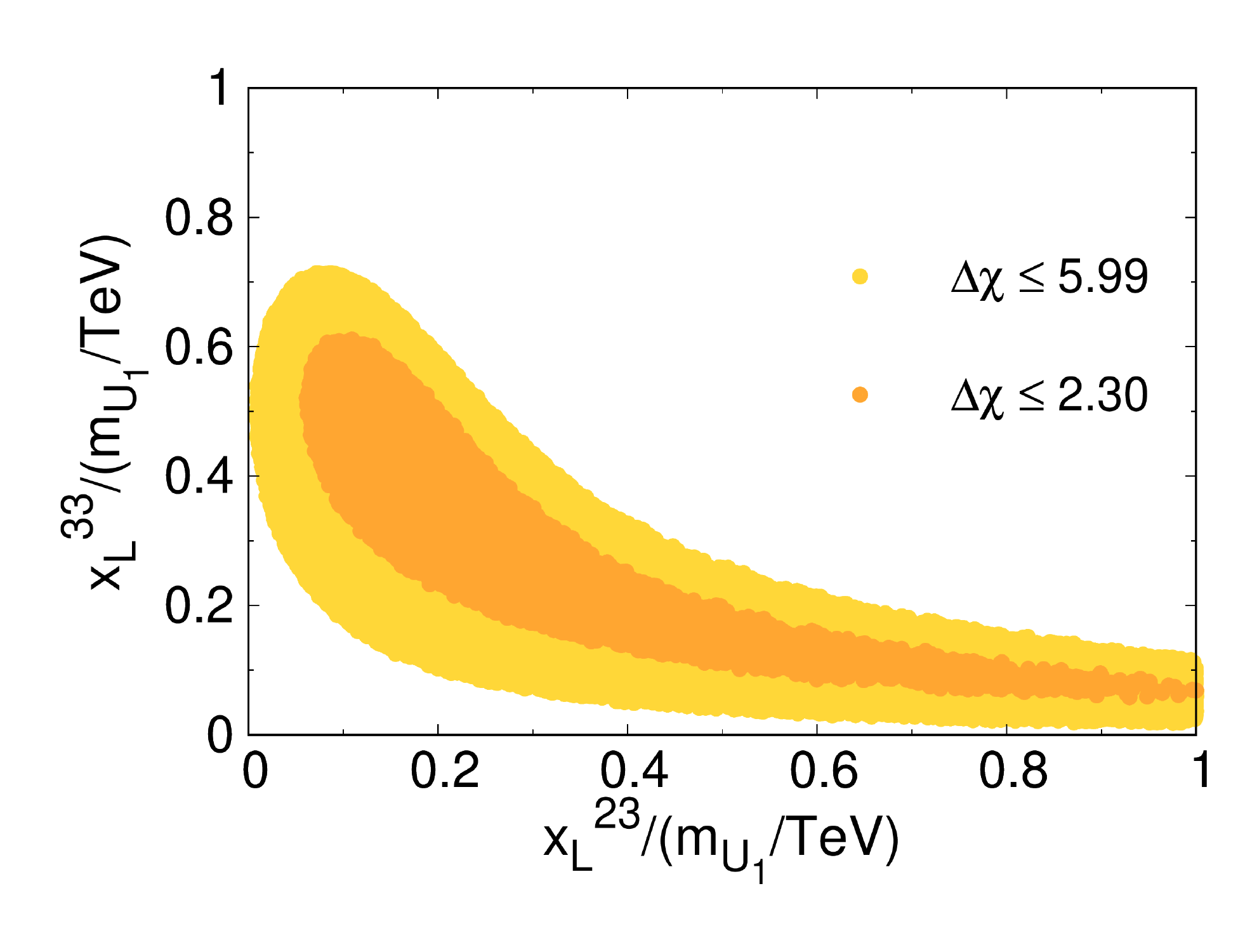}
		\includegraphics[height=1.5in,angle=0]{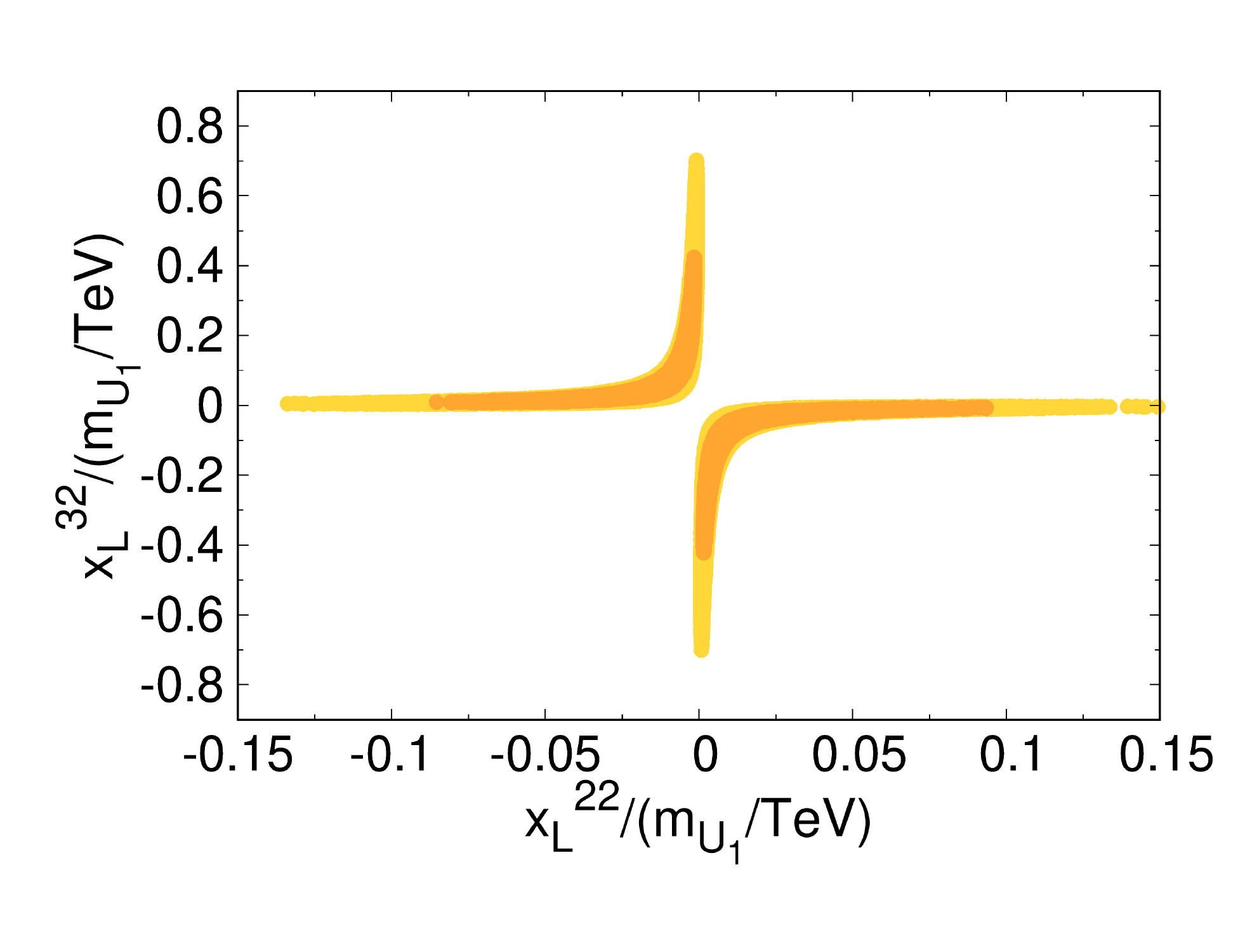}
		\includegraphics[height=1.5in,angle=0]{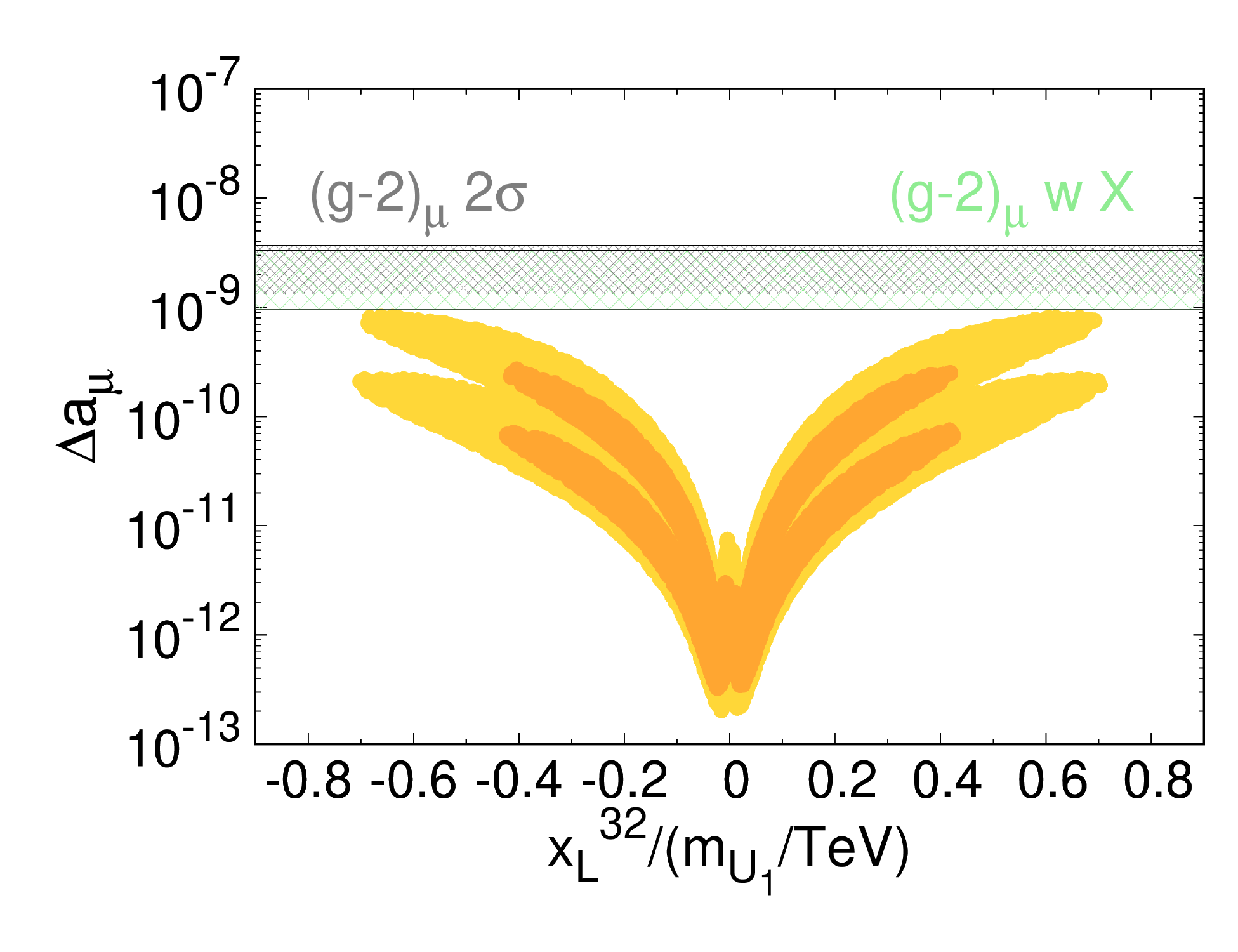}
		\includegraphics[height=1.5in,angle=0]{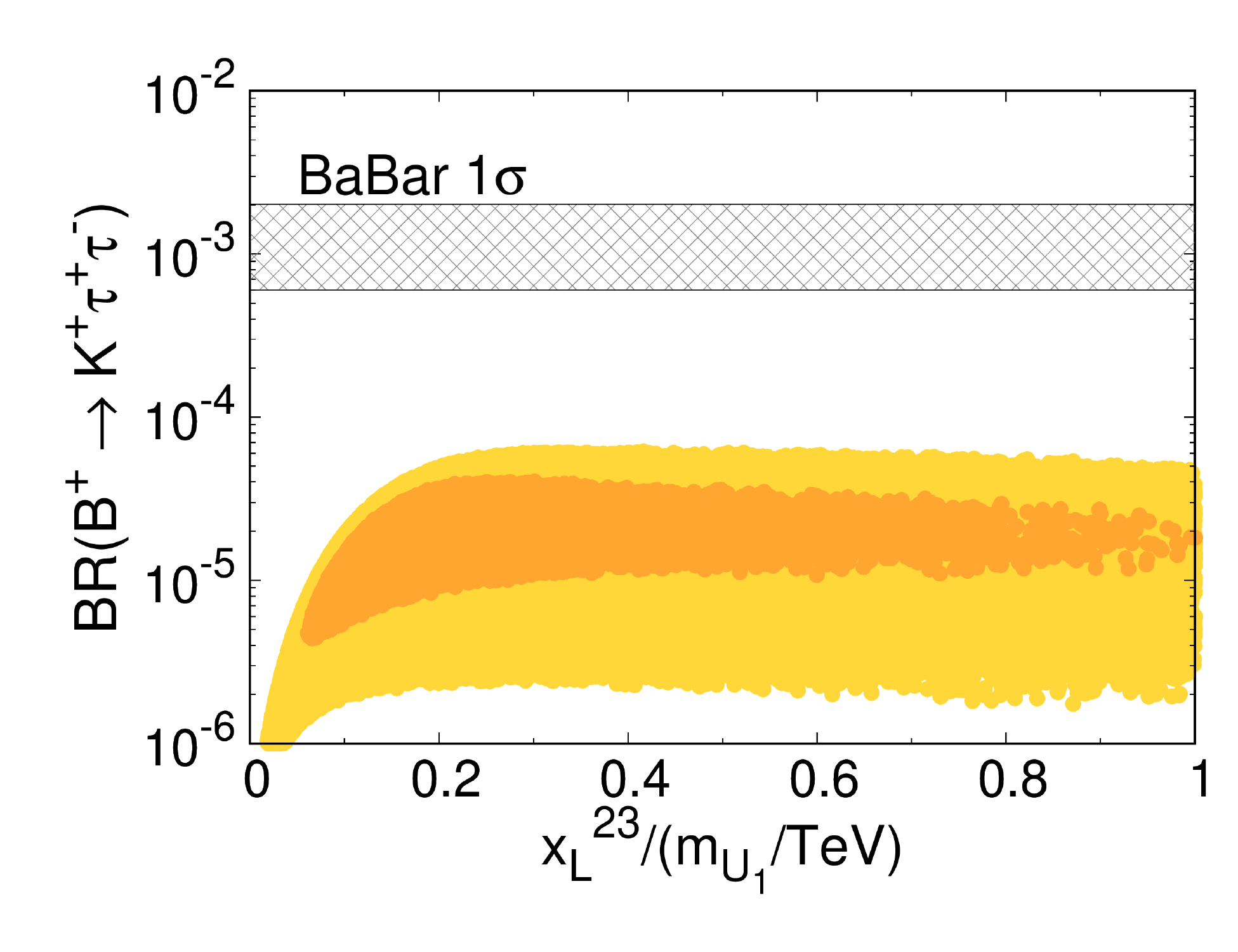}
		\includegraphics[height=1.5in,angle=0]{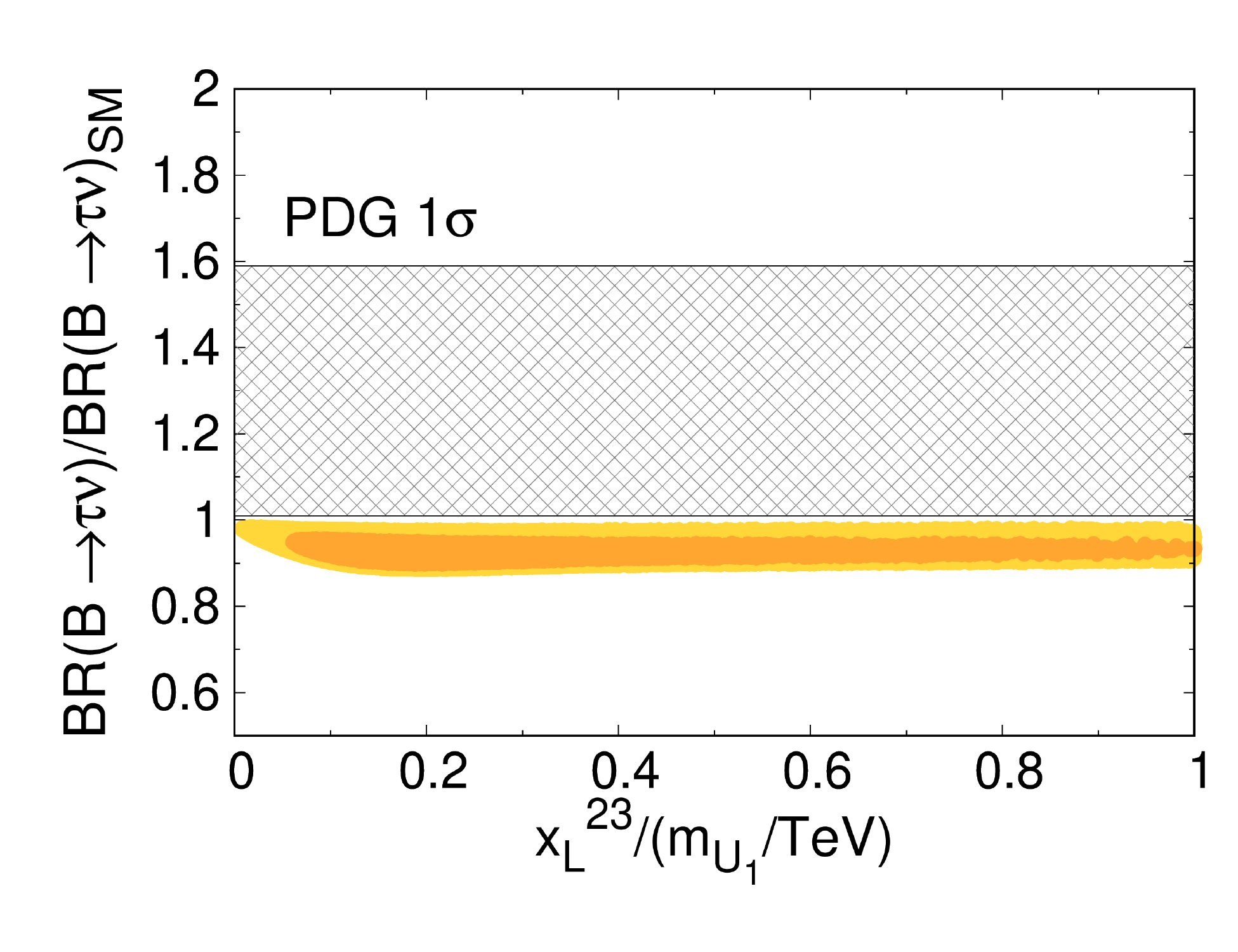}
		\includegraphics[height=1.5in,angle=0]{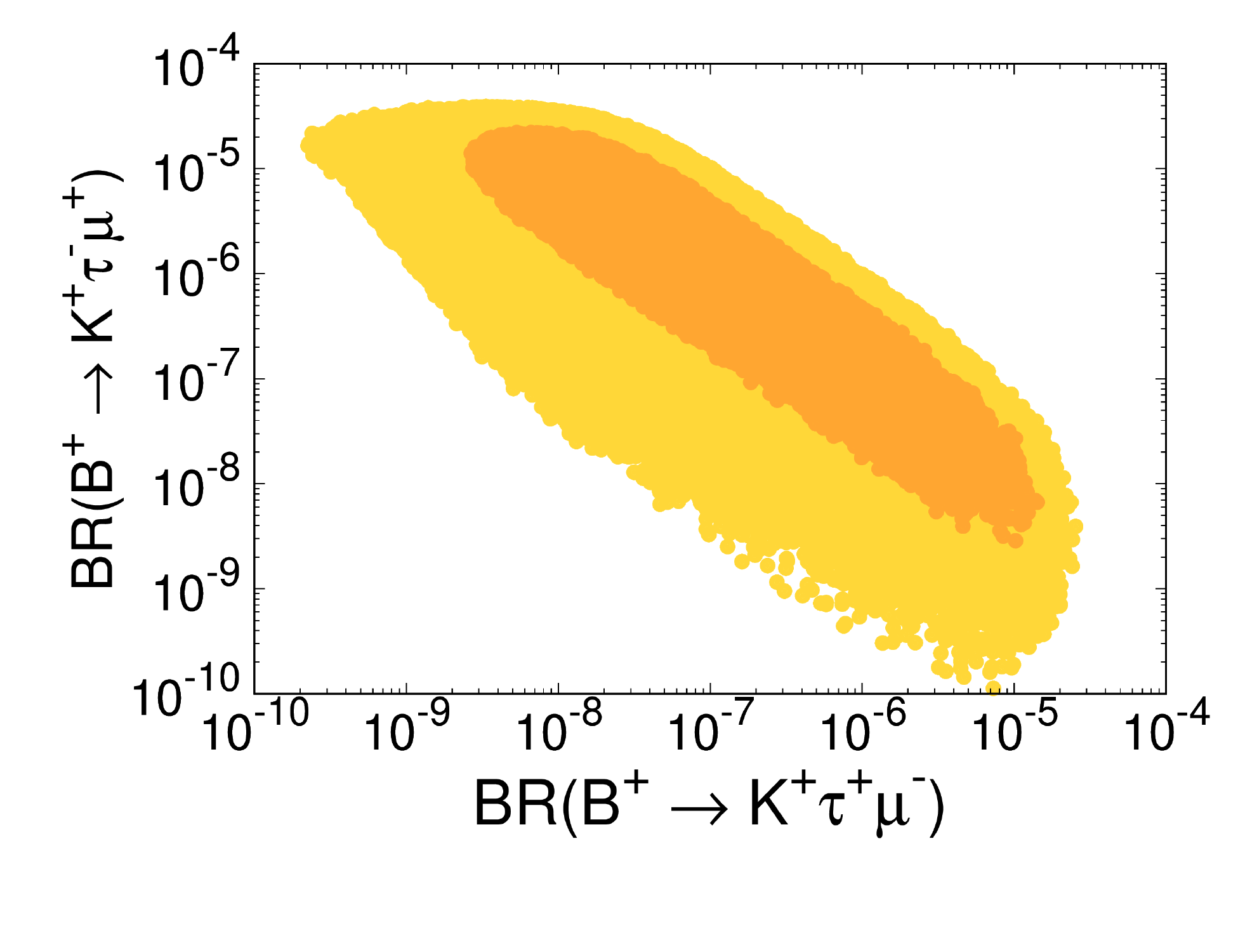}
		\includegraphics[height=1.5in,angle=0]{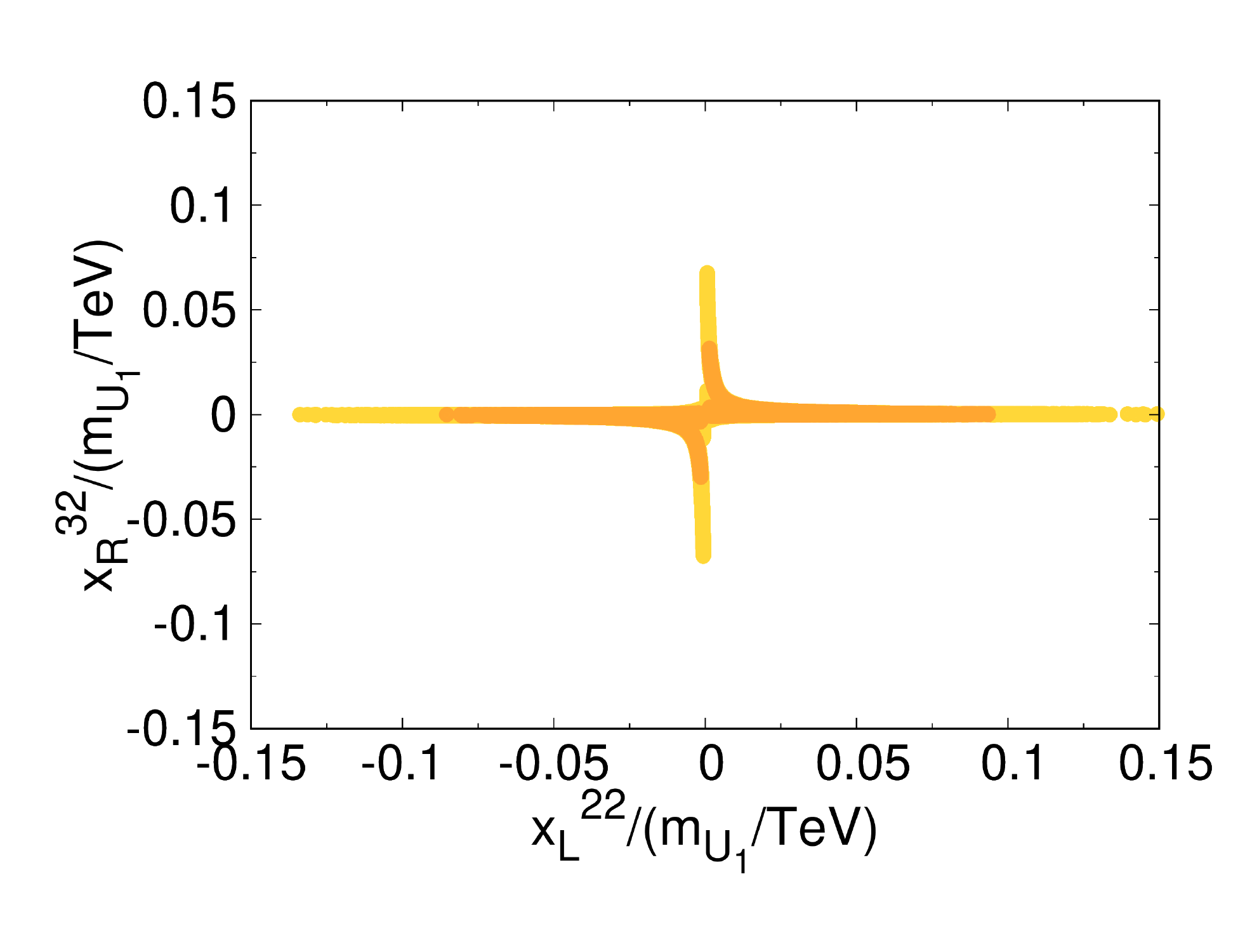}
		\includegraphics[height=1.5in,angle=0]{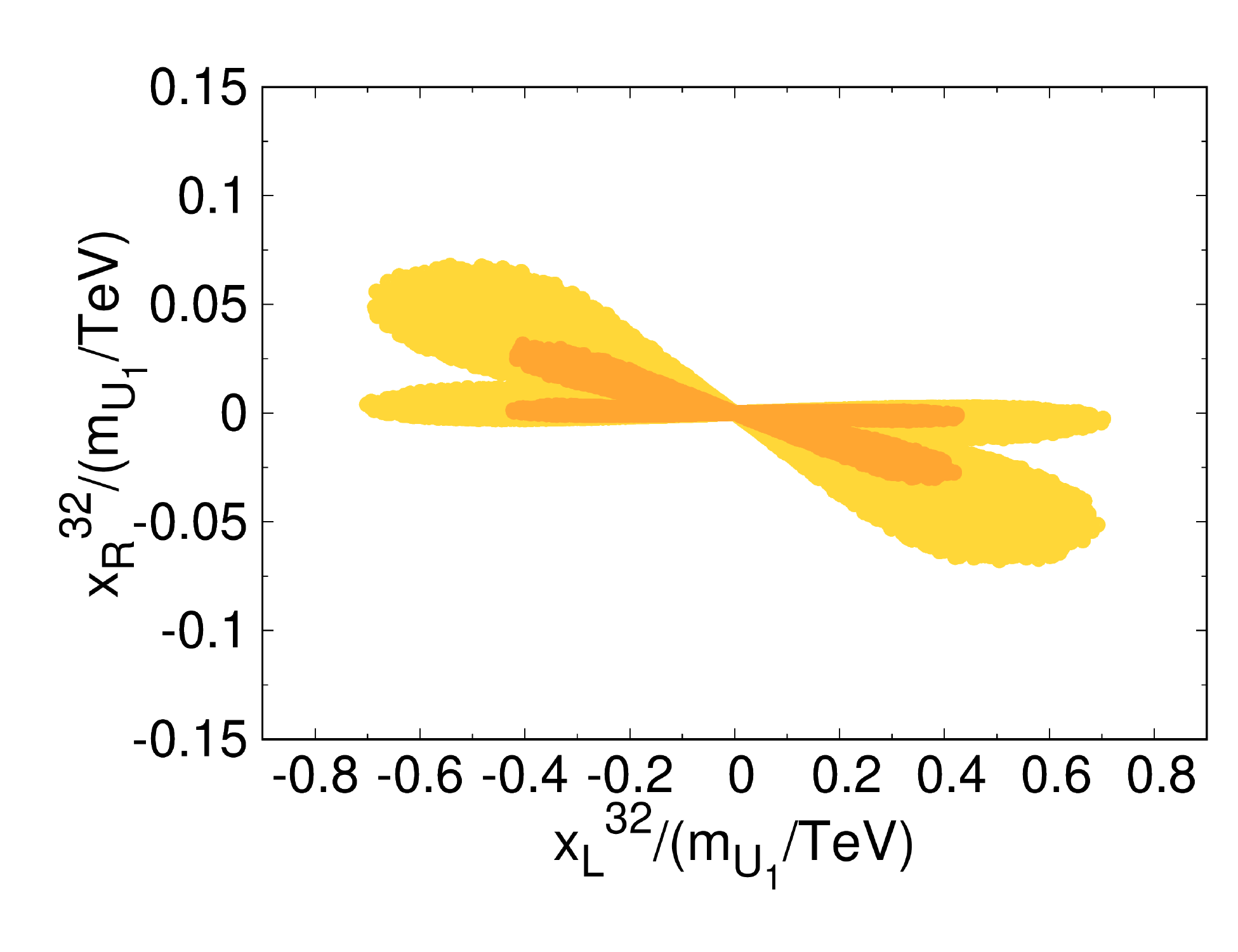}
		\includegraphics[height=1.5in,angle=0]{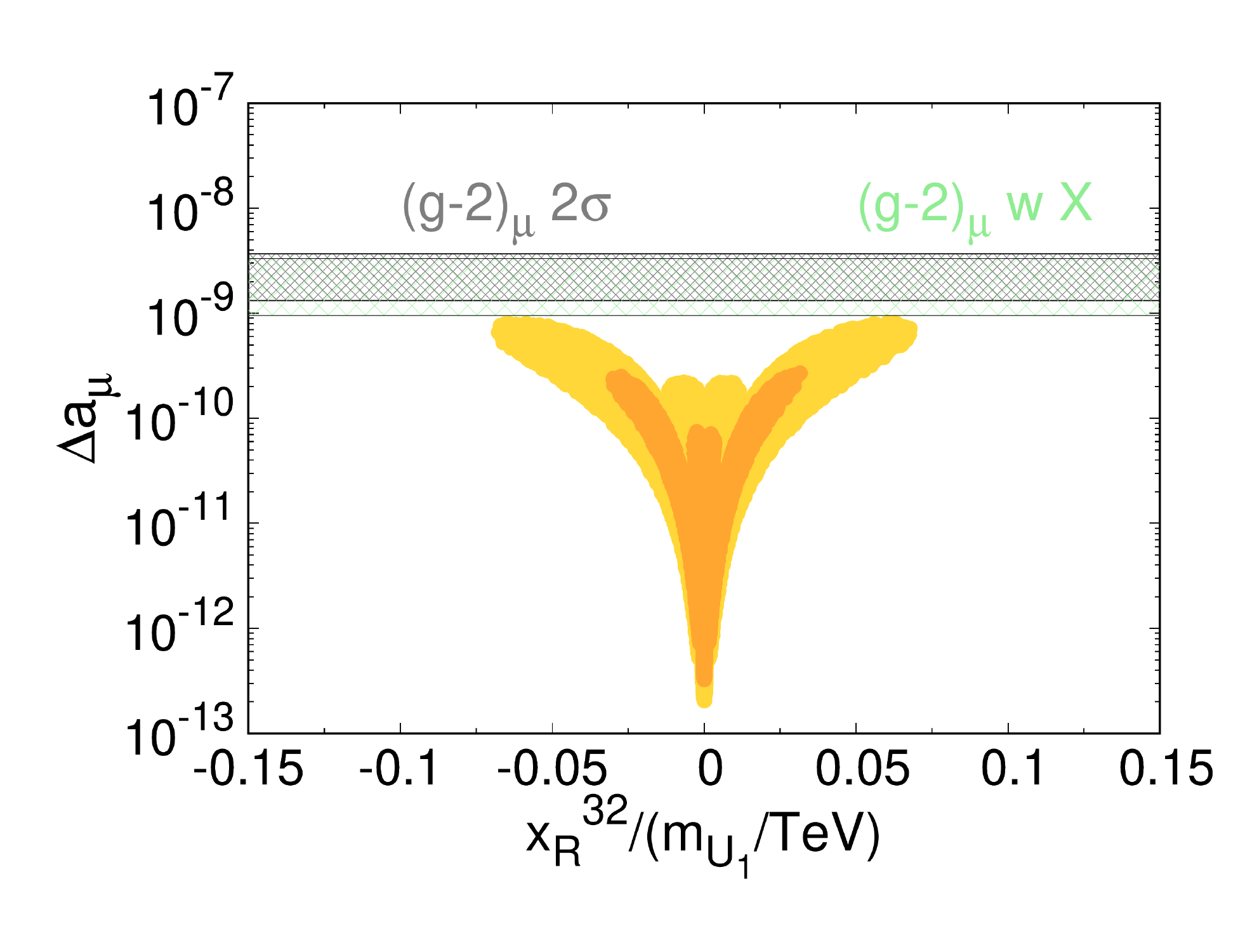}
		\includegraphics[height=1.5in,angle=0]{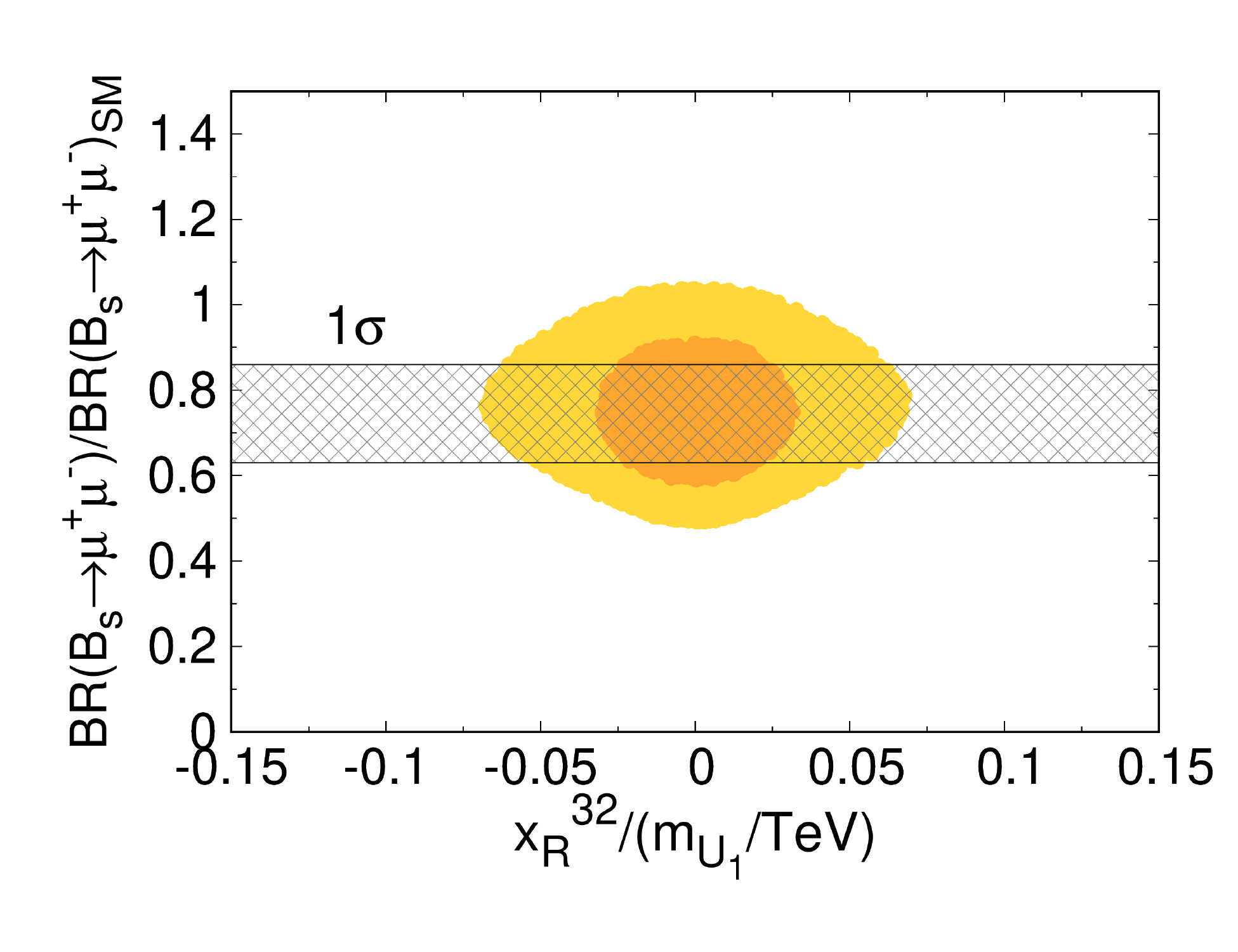}
		\caption{\small \label{fig:scan2}
			{\bf Scan-2:} The $1\sigma$ ($2\sigma$) region satisfies 
			$\Delta \chi^2 \equiv \chi^2 - \chi^2_{\rm min,2}\leq 2.30$ ($5.99$), 
			and $\chi^2_{\rm min,2}=9.06$.
			Here we fix $m_{U_1}=2$ TeV.
			The green hatched regions show the effect of a muon-philic $X$ vector boson with coupling $g_X=0.2$ and mass $m_X=100$ GeV, which endows $\Delta a_\mu|_X=38\times 10^{-11}$.
		}
	\end{figure}

\begin{figure}[ht]
		\centering
	    \includegraphics[height=1.5in,angle=0]{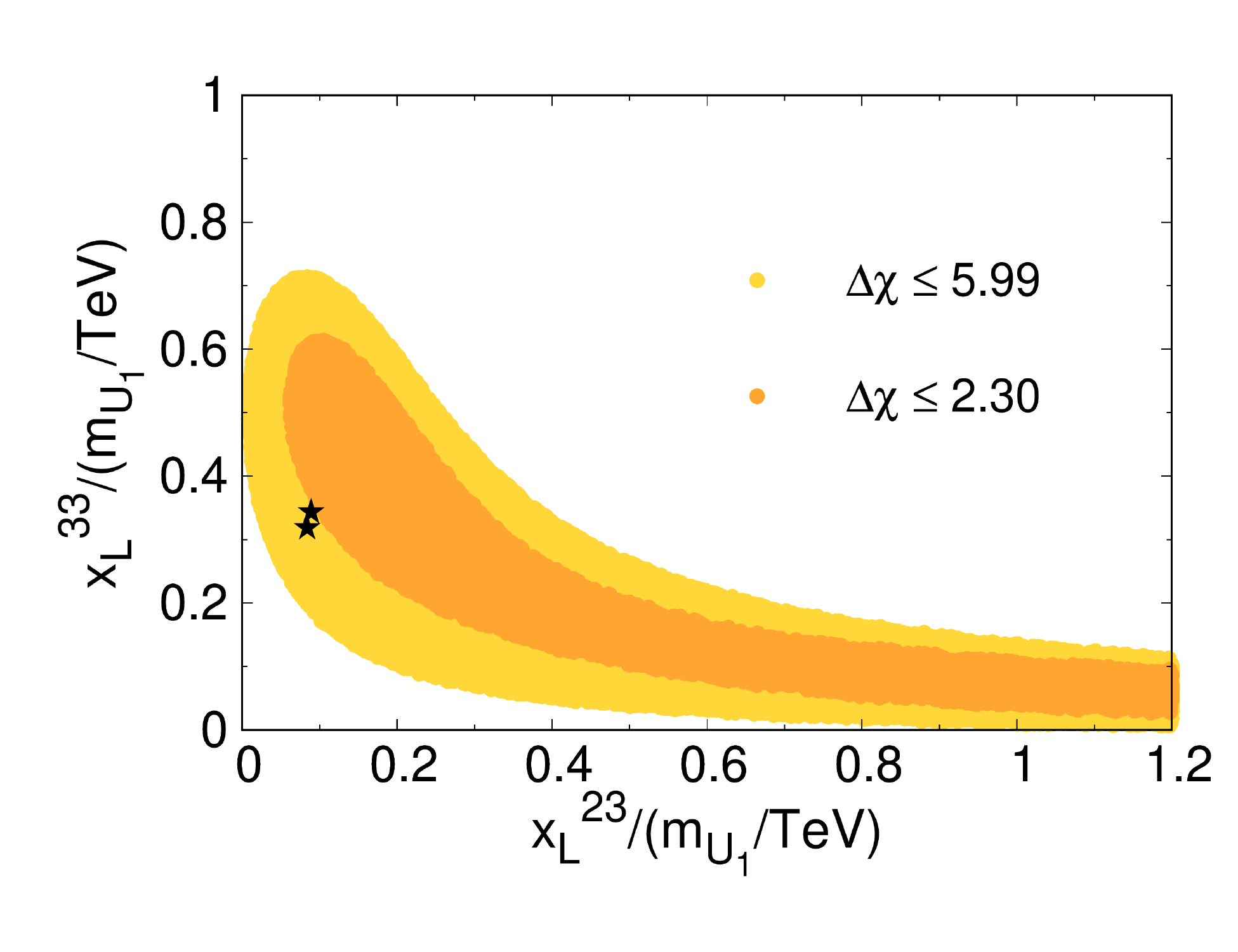}
		\includegraphics[height=1.5in,angle=0]{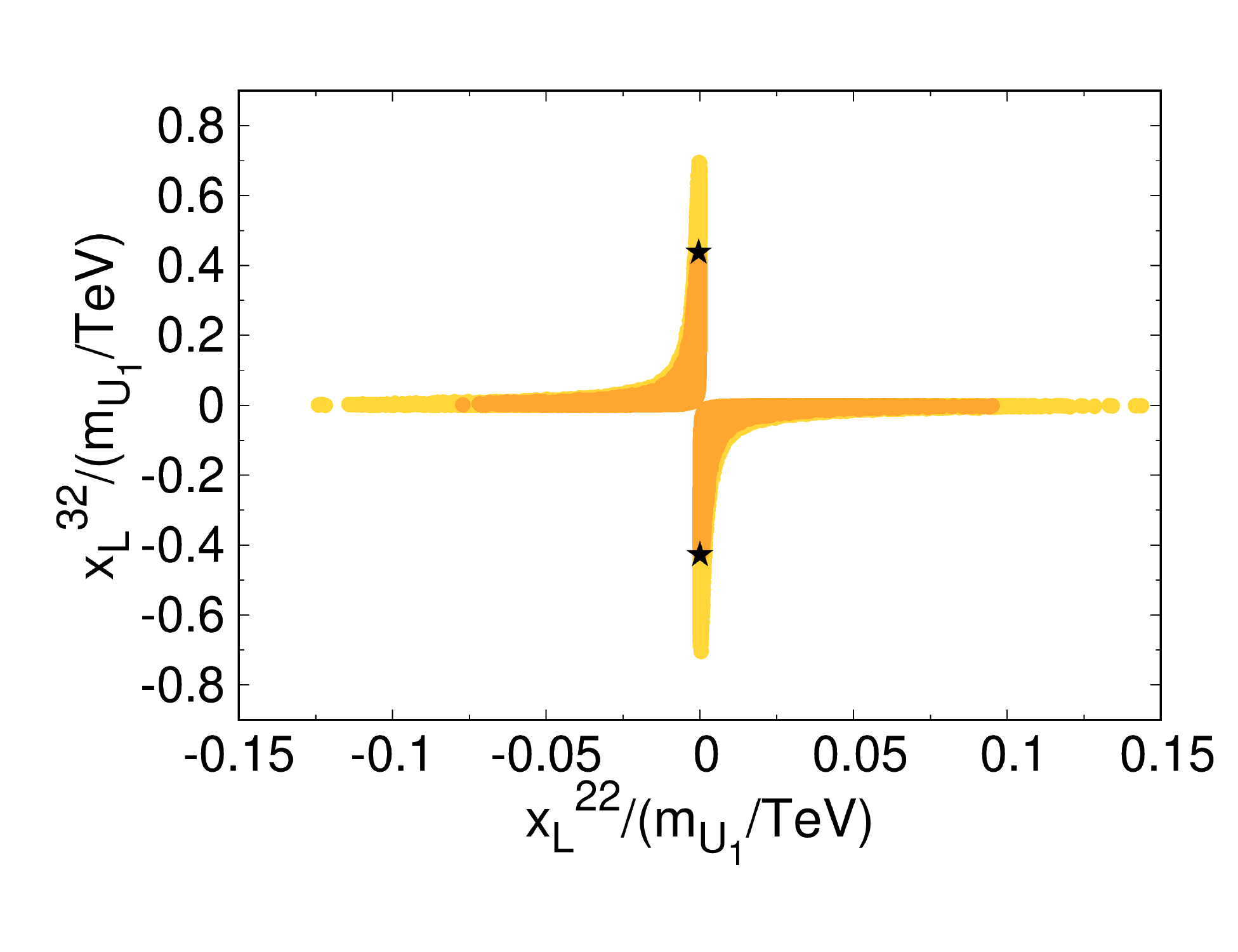}
		\includegraphics[height=1.5in,angle=0]{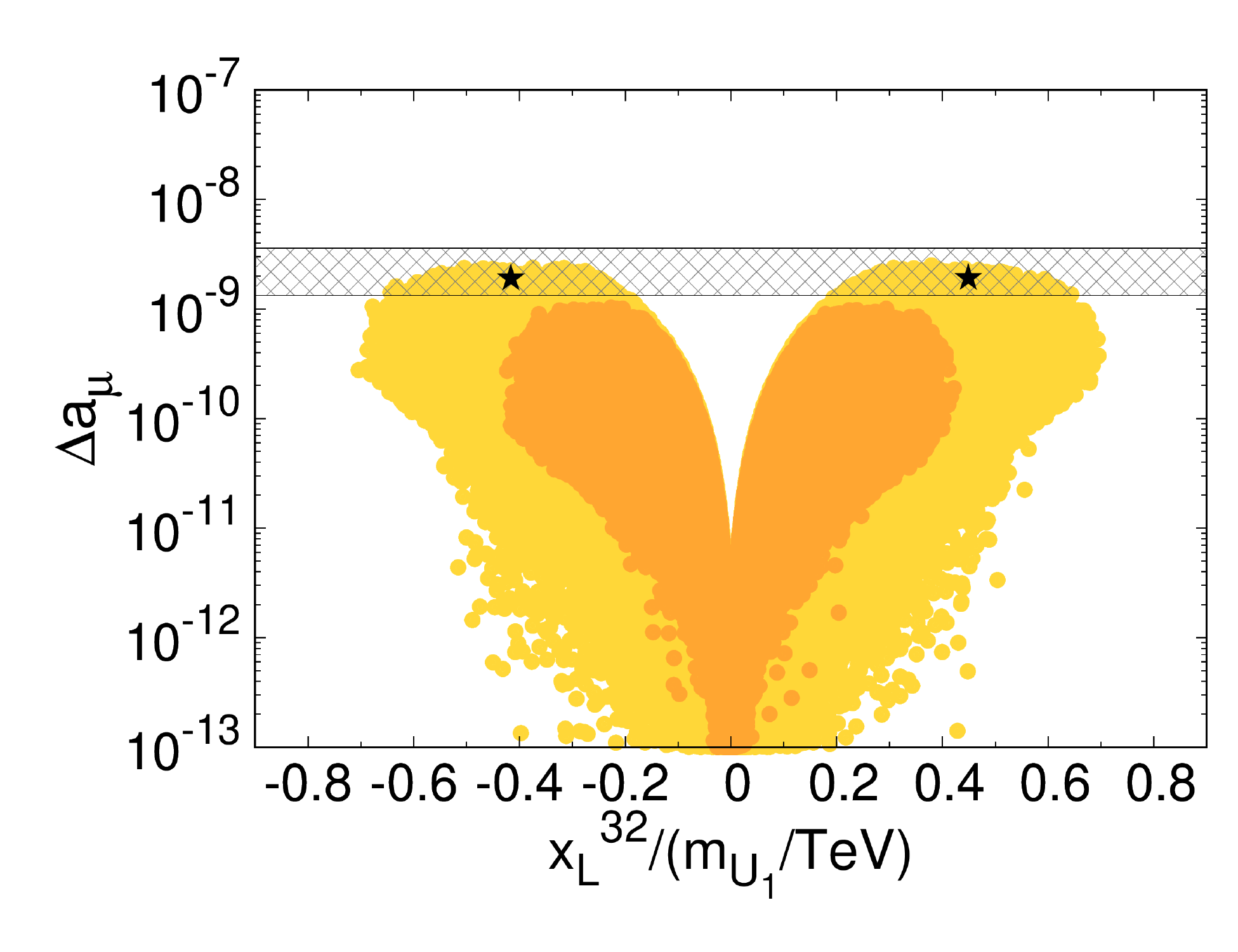}
		\includegraphics[height=1.5in,angle=0]{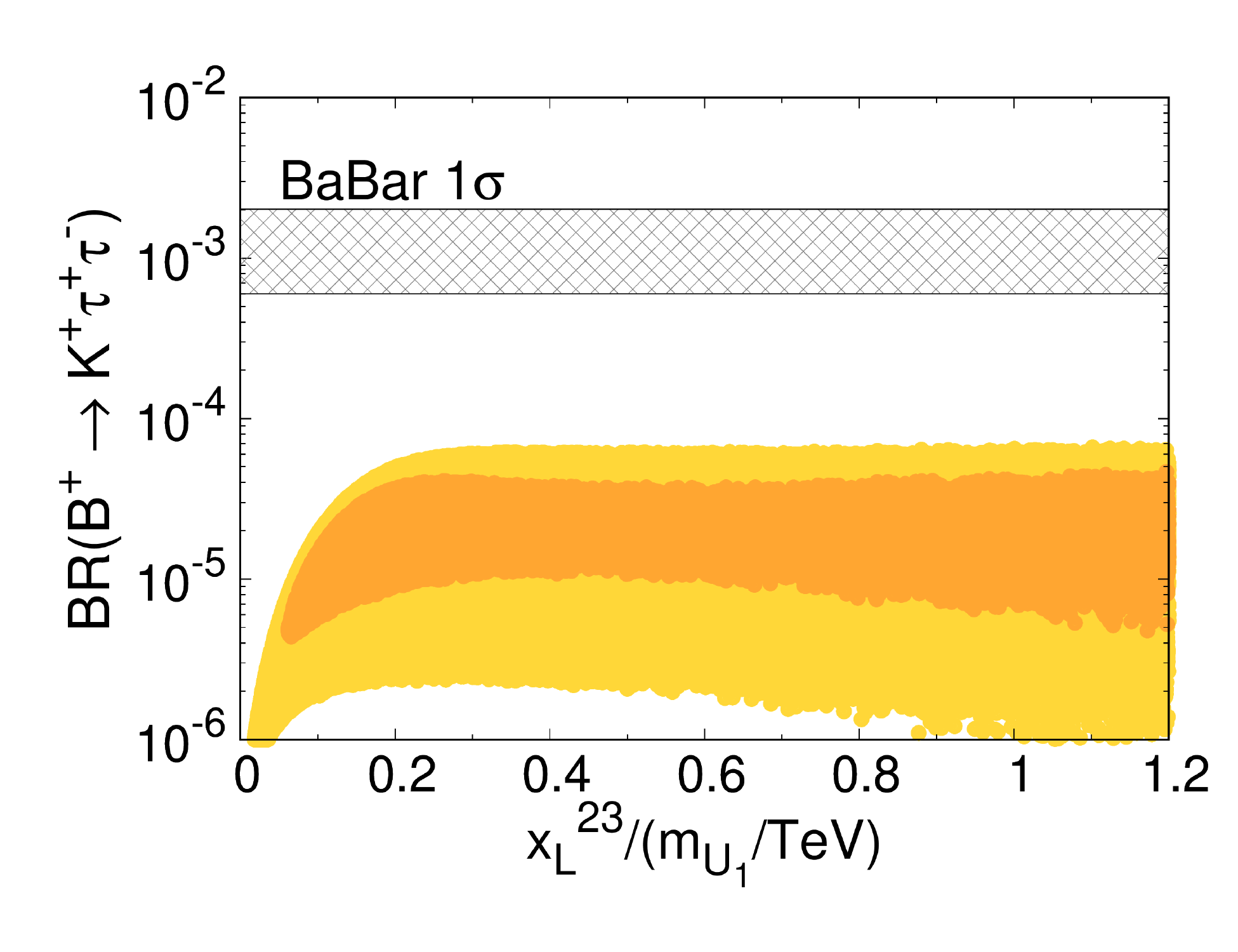}
		\includegraphics[height=1.5in,angle=0]{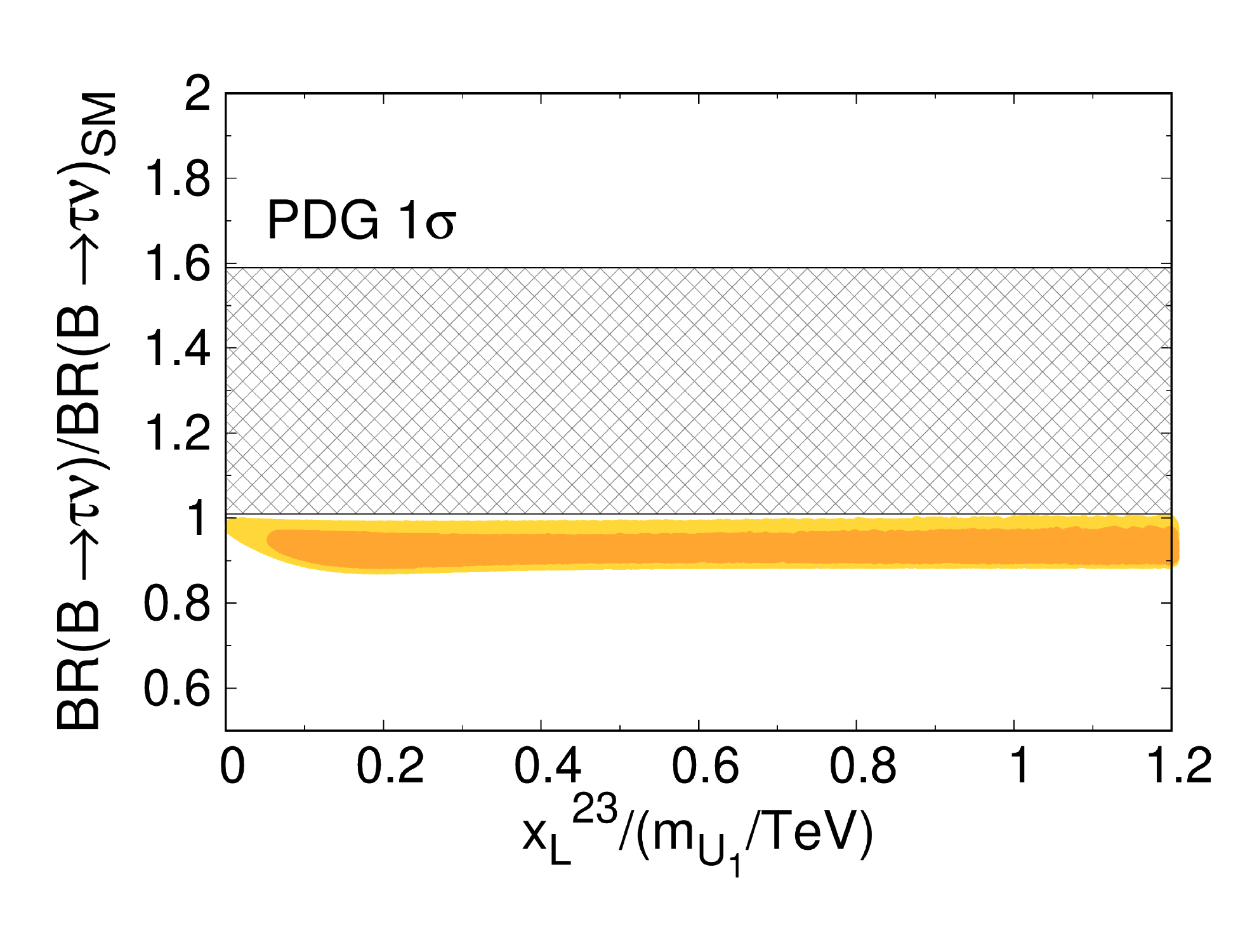}
		\includegraphics[height=1.5in,angle=0]{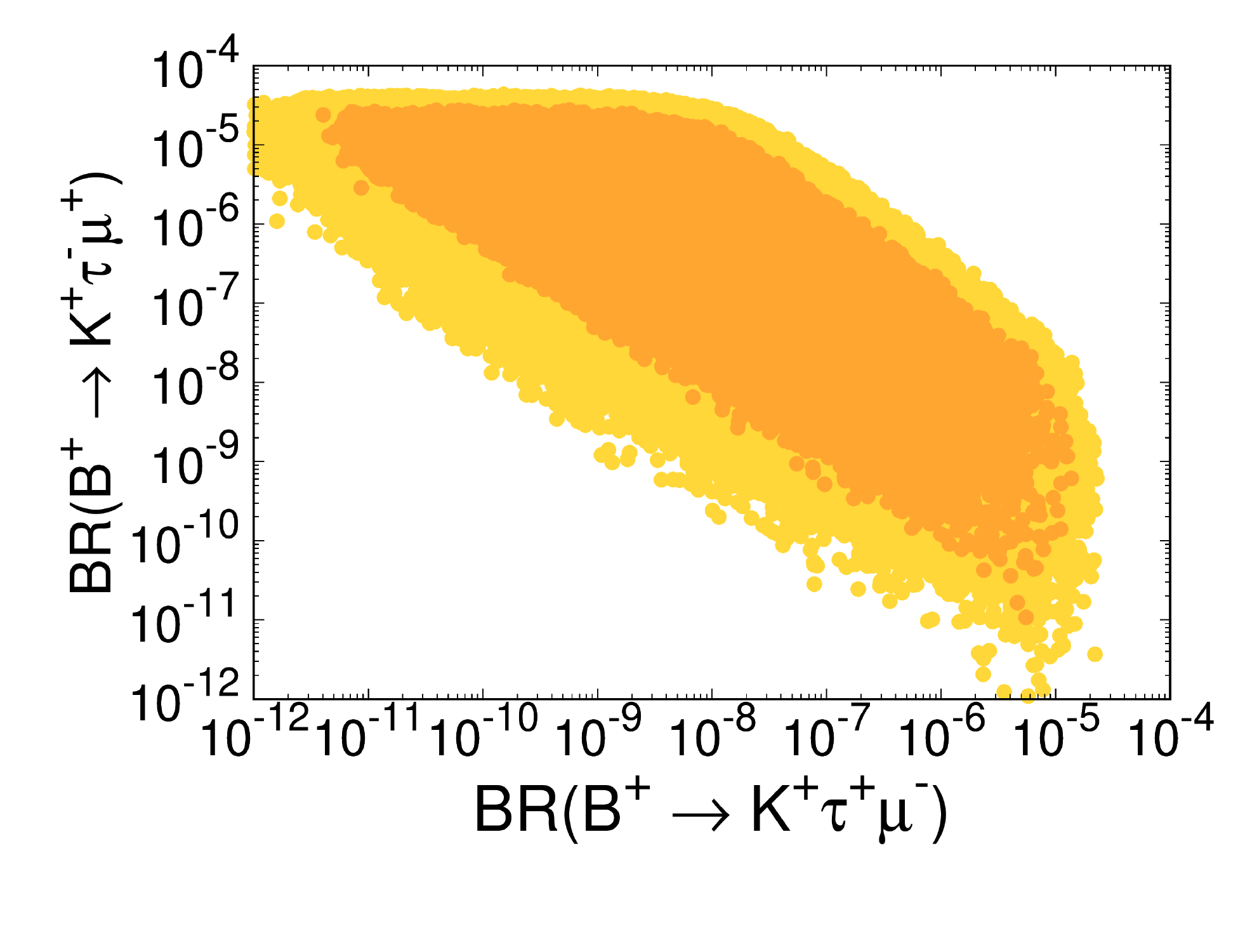}
		\includegraphics[height=1.5in,angle=0]{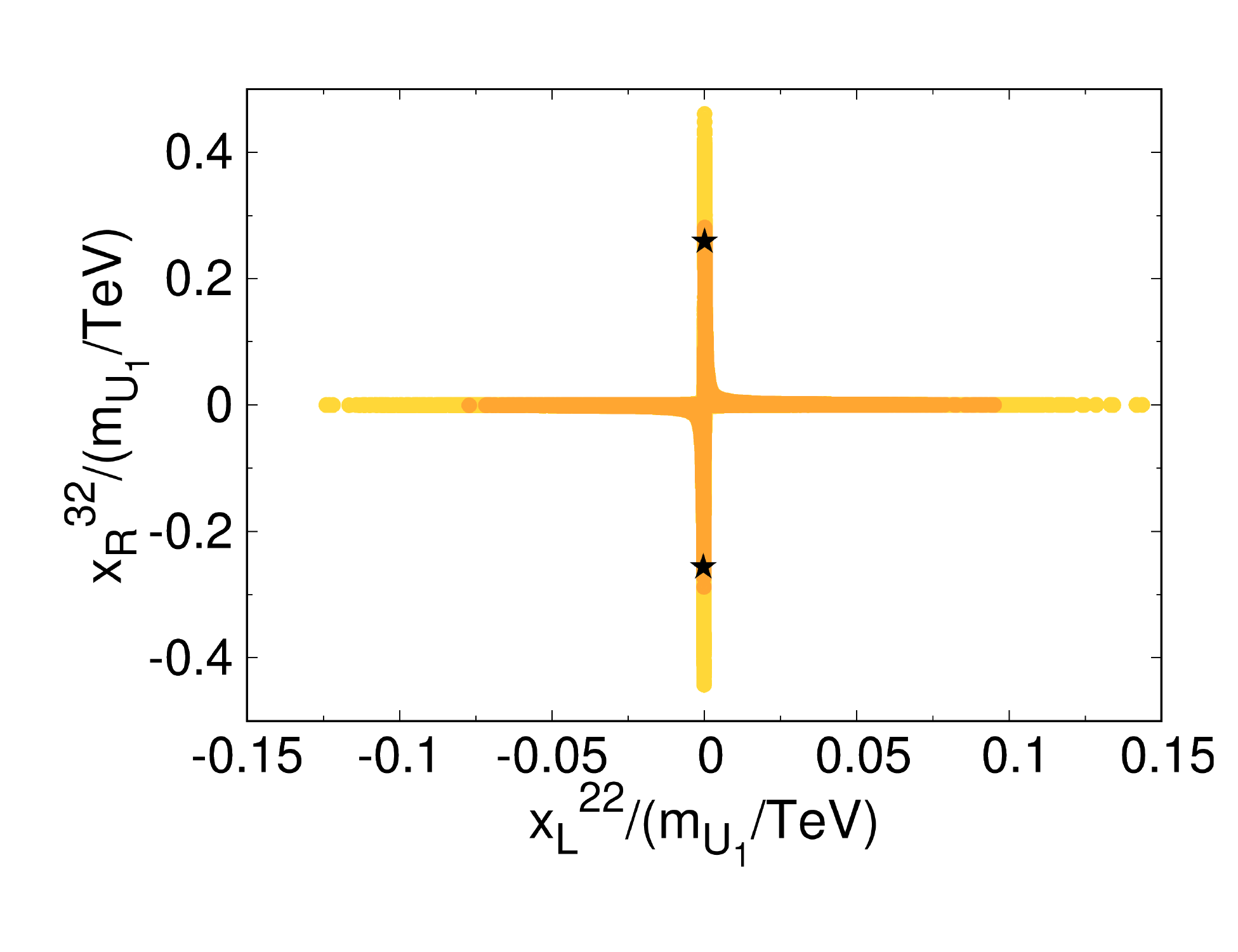}
		\includegraphics[height=1.5in,angle=0]{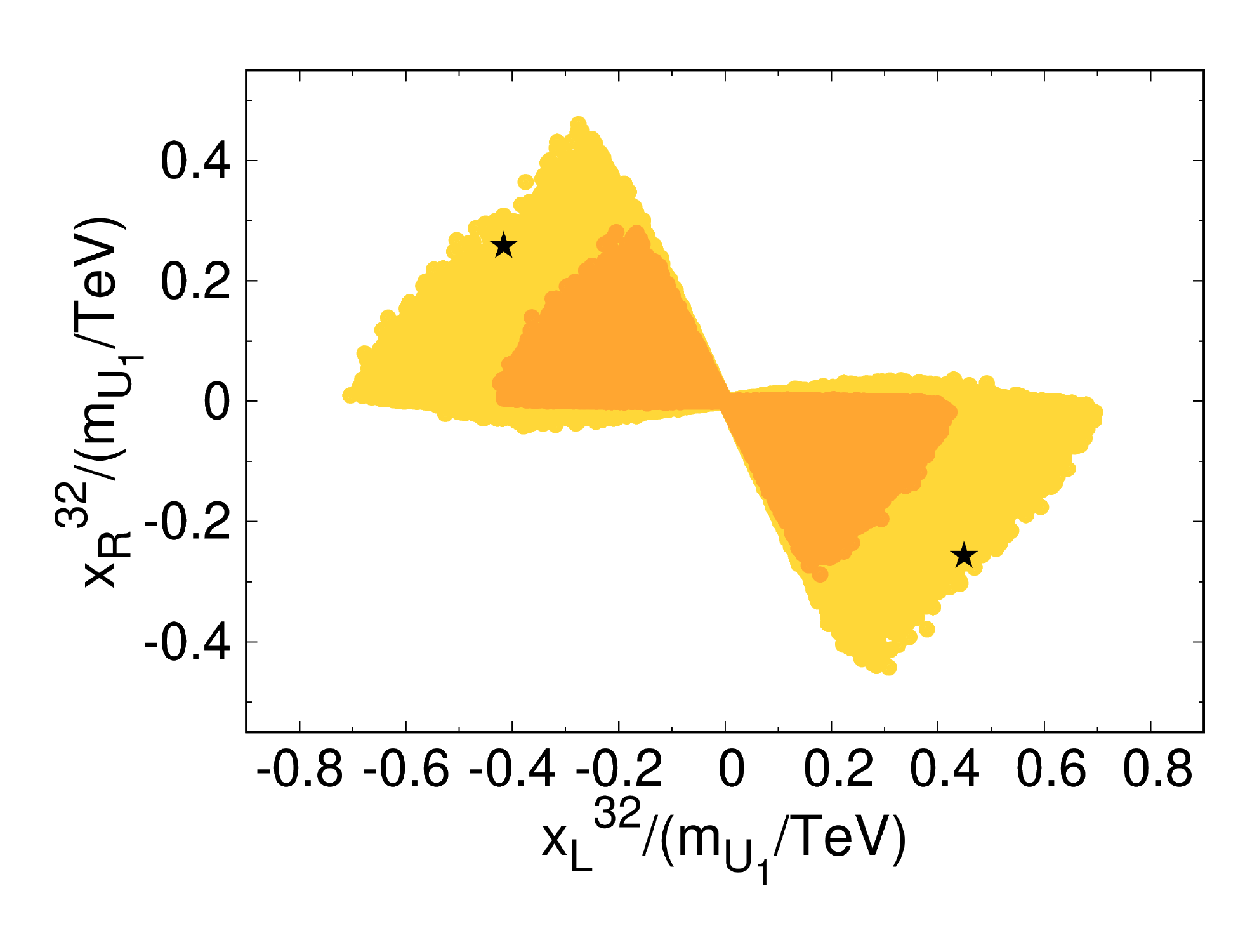}
		\includegraphics[height=1.5in,angle=0]{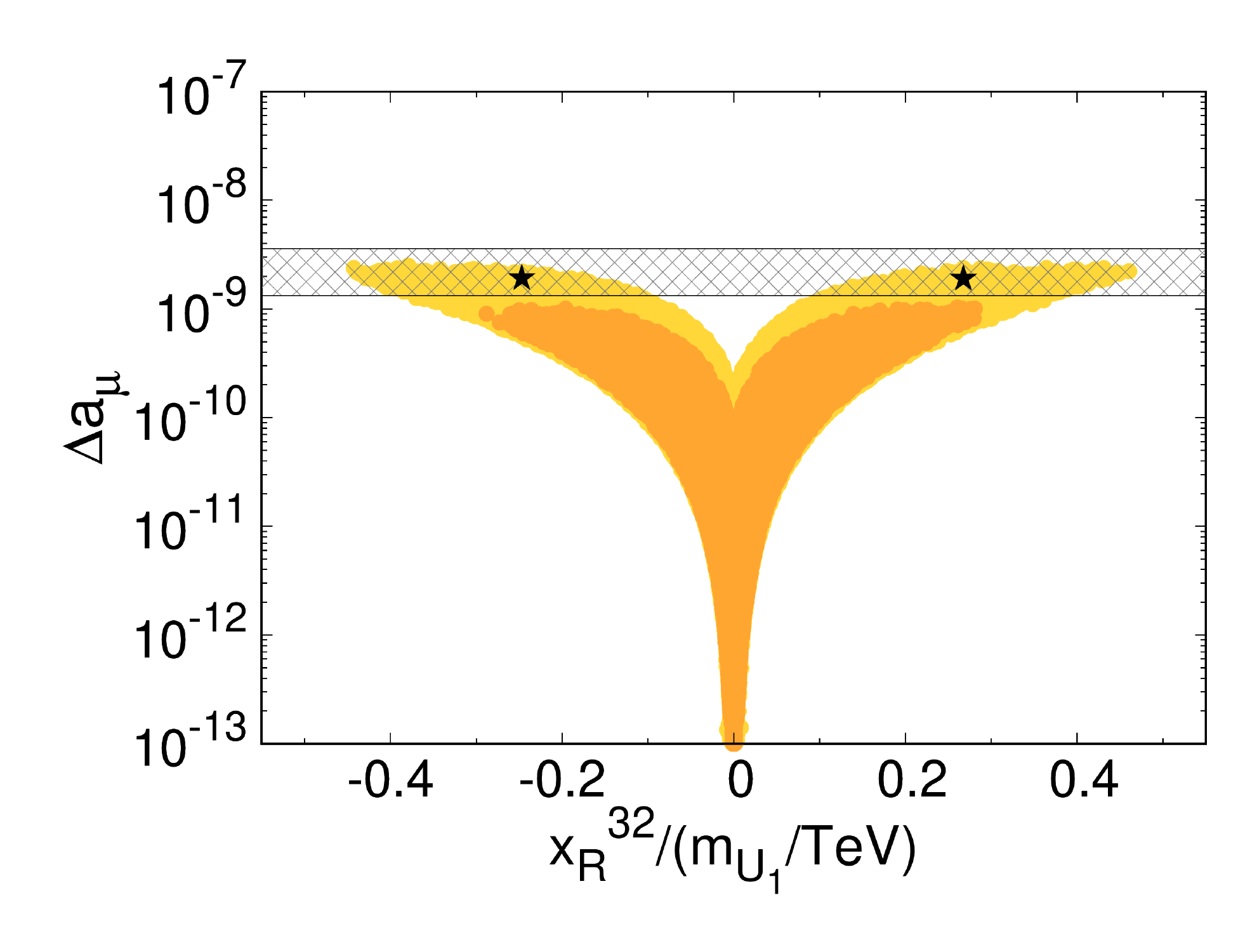}
		\includegraphics[height=1.5in,angle=0]{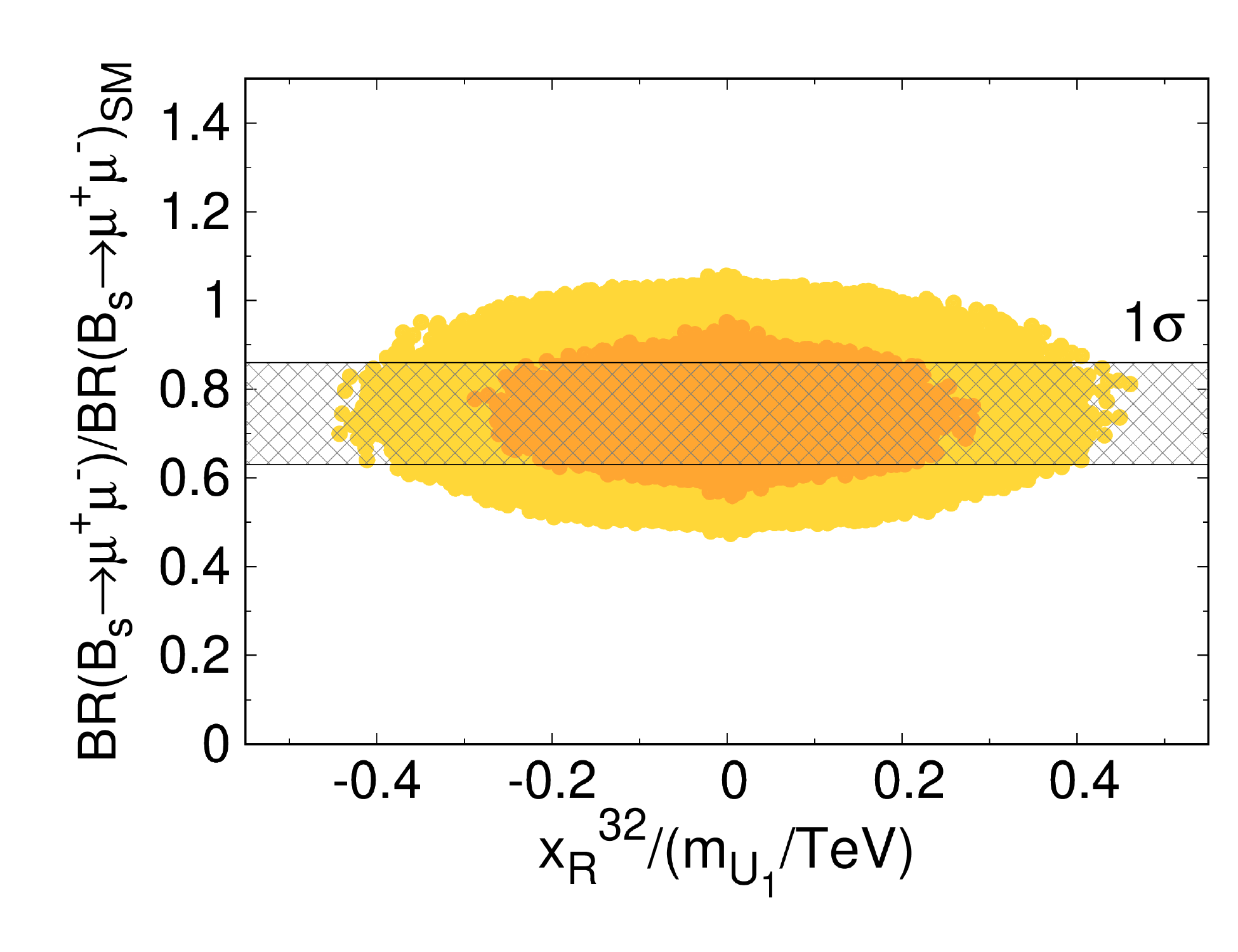}
		\caption{\small \label{fig:scan3}
{\bf Scan-3:} Same as Fig.~\ref{fig:scan2}, but includes the contribution of the additional $U(1)_{B_3-L_2}$ $X$ boson. 
Here we vary $0.01 \leq g_X \leq 0.05$, meanwhile fix $m_{U_1}=2$ TeV and $m_X=100$ GeV.
		}
\end{figure}

		We perform the $\chi^2$ parameter scanning, 
with the values of all 17 observables constituting the $\chi^2$ 
listed in Table \ref{tab1},
		which includes anomalies of $R_{K^{(*)}}$, $R_{D^{(*)}}$, 
		constraints from other $B$-meson decay channels, and constraints from $\tau$ decays.
Under the null hypothesis (SM only), we have $\chi^2_{\rm SM}=26.0$ along with $p_{\rm SM} = 0.074$ where $p_{\rm SM}$ is the $P$-value of the null hypothesis. We compare this result with the following three scenarios:
	
	\scp{
	\begin{itemize}
		\item {\bf scan-1}:  $P_{\text{scan-1}}=(x^{22}_L$, $x^{23}_L$, $x^{32}_L$, $x^{33}_L)$, 
		with $m_{U_1}=2,5~{\rm TeV}$.\\
		Results are in Fig.~\ref{fig:scan1}.
		\item {\bf scan-2}:  $P_{\text{scan-2}}=P_{\text{scan-1}} \oplus x^{32}_R$, 
		with $m_{U_1}=2~{\rm TeV}$.\\
		Results are in Fig.~\ref{fig:scan2}.
	    \item {\bf scan-3}: $P_{\text{scan-3}} =P_{\text{scan-2}} \oplus g_X$, with $m_{U_1}=2$ TeV, and  $m_X=100~{\rm GeV}$.\\
		Results are in Fig.~\ref{fig:scan3}.
	\end{itemize}
	}
		
		For {\bf scan-1}, we choose a set of relevant couplings, 
		$x^{22}_L$, $x^{23}_L$, $x^{32}_L$, $x^{33}_L$, 
		that is sufficient to explain the $B$-meson anomalies 
		and satisfies all the low-energy observables. 
		It gives rise to the best fit $\chi^2_{\rm min,1}=9.23$ and $p_1 = 0.755$   %$\chi^2_{\rm min,1}=9.18$ and $p_1 = 0.759$
		with $(x^{22}_L,x^{23}_L,x^{32}_L,x^{33}_L)=(7.90\times 10^{-2},0.328,-3.83\times 10^{-2},0.862)$ and $m_{U_1}=2~{\rm TeV}$.
		%with $(x^{22}_L,x^{23}_L,x^{32}_L,x^{33}_L)=(5.42\times 10^{-2},0.279,-5.05\times 10^{-2},0.947)$ and $m_{U_1}=2~{\rm TeV}$.
		%
		Figure \ref{fig:scan1} shows the $1\sigma$ region for the {\bf scan-1}.
		The favored region of the $(x^{23}_L,x^{33}_L)$ plane, shown in the upper-left panel of Fig.~\ref{fig:scan1}, is mainly determined 
		by the $R_{D^{(*)}}$, meanwhile the $R_{K^{(*)}}$ dictates the favored region of the $(x^{22}_L,x^{32}_L)$ plane.
		There is contribution to $\Delta a_\mu$ from the LH couplings, as shown in the upper-right panels
		of Fig.~\ref{fig:scan1}. However, it is not large enough to explain the recent measurment from Fermilab,
		\yjk{thus providing} the motivation to turn on the RH coupling in the {\bf scan-2}.

		In the {\bf scan-2} ($\chi^2_{\rm min,2}=9.06$ and $p_2 = 0.698$),  %($\chi^2_{\rm min,2}=9.18$ and $p_2 = 0.687$), 
		according to Eq.(\ref{eq:g_2}), we found the most efficient way 
		to enhance $\Delta a_\mu$ is to use $x^{32}_R$, 
		due to the fact that the multiplicity of $(x^{32}_L x^{32}_R)$ has milder mass suppression factor
		$(m_b m_\mu/m_{U_1})$ than that of pure LH couplings or RH couplings. 
		In the $(x^{32}_L,\Delta a_\mu)$ and 
		$(x^{32}_R,\Delta a_\mu)$ planes of Fig.~\ref{fig:scan2},
		they show that the value of $\Delta a_\mu$ can be increased 
		by more than an order of magnitude in contrast to the
		{\bf scan-1}.
		However, it still cannot explain the BNL and FNAL results without additional contribution. 
		For example, incorporating a muon-philic vector boson $X$ with coupling $g_X=0.2$
		and mass $m_X=100$ GeV, 
		which is consistent with current experimental bounds, 
		endows $\Delta a_\mu|_X=38\times 10^{-11}$ shown 
		by the green hatched regions overlaping 
		with $U_1$ $2\sigma$-allowed region 
		in Fig.~\ref{fig:scan2}.

   {In the {\bf scan-3}, we include a specific $U(1)_{B_3-L_2}$ $X$ boson 
    in additiona to the $U_1$ leptoquark framework, 
    and demonstrate that under this framework it can alleviate 
    the $(g-2)_\mu$ and $B$-physics tensions within $2\sigma$.
    Not only the $(g-2)_\mu$, but this $X$ boson also contributes to both
    $\Delta C_9$ (see Eq.(\ref{eq:C9_X})) and $B_s \to \mu \mu$ 
    (see Eq.(\ref{eq:Bs_mumu})). In particular, we vary the coupling 
    $0.01 \leq g_X \leq 0.05$, and fix $m_X=100~{\rm GeV}$, 
    which { is partially allowed} under current experimental constraints from neutrino-trident~\cite{Altmannshofer:2014pba}, 
    $B^0_s-\bar{B}^0_s$ mixing~\cite{DiLuzio:2019jyq}, $K_L \to \mu^+\mu^-$~\cite{Buras:1997fb}, 
    ATLAS~\cite{Pandolfi:2017lpu}, and CMS~\cite{CMS:2017moi,CMS:2021ctt}. 

    {
    It is worth to mention the reliable range of $X$ boson parameters consistent with the measurement of i) $B_s^0 - \bar{B}_s^0$ mixing ($\Delta M_s$), ii) $\text{Br}(K_L \to \mu^+ \mu^-$) and iii) $K^0 - \bar{K}^0$ mixing parameters. The mass difference $\Delta M_s$ has been precisely measured by CDF2, LHCb and CMS collaborations \cite{HFLAV:2016hnz,HFLAV:2022pwe}, and its theoretical predictions has been improved by {developed} sum rules and lattice calculations. We use the weighted average of the latest results given in Ref.~\cite{DiLuzio:2019jyq} as our $\Delta M_s^{\rm SM}$. In the presence of $X$ boson, new physics contribution to $\Delta M_s$ is approximated by $\frac{\Delta M_s^{\rm SM+NP}}{\Delta M_s^{\rm NP}} \approx \Bigl | 1 + \frac{1}{360} \Bigl ( \frac{\delta C_9^\mu}{-0.53} \Bigr )^2 \Bigl ( \frac{m_X/g_X}{1 \text{ TeV}} \Bigr )^2 \Bigr |$ \cite{DiLuzio:2019jyq}. Putting $\delta C_9^\mu (=-\delta C_{10}^\mu)=-0.40\pm0.12$ \cite{Aebischer:2019mlg,Alguero:2019ptt,Cornella:2019hct} into the expression and requiring $\Delta M_s = (1.04_{-0.07}^{+0.04}) \Delta M_s^{\rm exp}$, the lower limit on the coupling is $g_X > 1.01 \times 10^{-2}$ within 1$\sigma$ for $m_X = 100$ GeV.\footnote{Using the global fit $\delta C_9^\mu (=-\delta C_{10}^\mu) = -0.39 \pm 0.07$ based on Moriond 2021 result \cite{Altmannshofer:2021qrr}, the lower limit on the coupling is $g_X > 1.15 \times 10^{-2}$ within 1$\sigma$ for $m_X = 100$ GeV.} For Br($K_L \to \mu^+ \mu^-$), requiring that new physics contribution to dimension-six $\Delta F = 1$ operator $(\bar{s}_L \gamma^\mu d_L)(\bar{\mu} \gamma_\mu \mu)$ is smaller than SM contribution \cite{Buras:1997fb, DAmbrosio:2022kvb}, we impose the upper limit on the coupling $g_X < 4.61 \times 10^{-2}$ for $m_X = 100$ GeV. Focusing on the upper limit on the short-distance contribution Br($K_L \to \mu^+ \mu^-)_{\rm SD}$, a dedicate analysis gives $g_X |_{m_X = 100 \text{ GeV}} < 3.55 \times 10^{-2}$ \cite{Endo:2016tnu}. For the neutral Kaon mixing, the most stringent bound comes from the measurement of $\epsilon_K$. Following the approach in Ref.~\cite{Endo:2016tnu}, we get the allowed range on the coupling as $-4.13 \times 10^{-2} < g_X < 4.24 \times 10^{-2}$ for $m_X = 100$ GeV. We assume the left-handed coupling to $X$ boson is dominant for flavor-violating vertices in the hadronic sector.
    }
    
    Such $X$ boson has 
    negligible effect on $(g-2)_\mu$, but gives modest contribution to $\Delta C_9$.
    Due to the negative value of $V_{ts}\simeq -0.40$, 
    the $X$ boson trends to the observed value ($C^{\mu\mu}_9=-0.40 \pm 0.12$).
    Consequently, the $U_1$ leptoquark in conjunction with $X$ boson explain 
    the \yjk{$R_{K^{(*)}}$} anomaly, but the former solely contributes to the $(g-2)_\mu$. 
    Comparing to {\bf Scan-2}, the $\chi^2_{\rm min}=9.06$ at  %the $\chi^2_{\rm min}=9.18$ at 
    $(x^{22}_L,x^{23}_L,x^{32}_L,x^{33}_L,x^{32}_R,g_X)=(4.04\times 10^{-2},0.324,-9.30\times10^{-3},0.810,3.73\times10^{-4},1.47\times10^{-2})$    %$(x^{22}_L,x^{23}_L,x^{32}_L,x^{33}_L,x^{23}_R,g_X)=(0.051,0.29,-0.034,0.93,0.003,0.010)$
    is no further reduced by including the $X$ boson. 
    However, the $2\sigma$ chi-square regions (yellow regions in Fig.~\ref{fig:scan3}) 
    extend overlapping with $(g-2)_\mu$ $2\sigma$ region.
    Two representative points in the overlapped regions are
    \begin{eqnarray}
	(x^{22}_L,x^{23}_L,x^{32}_L,x^{33}_L,x^{32}_R,g_X)=\left\lbrace
	\begin{array}{c}
		(+3.28\times 10^{-4},0.167,-0.846,0.644,+0.520,1.32\times 10^{-2})\,, \nonumber \\ [2mm]
		%(+3.3\times 10^{-4},0.17,-0.85,0.64,+0.52,0.013)\,, \nonumber \\ [2mm]
		%
		(-3.08\times 10^{-4},0.227,+0.831,0.668,-0.554,1.44\times 10^{-2})\,, \nonumber
		%(-3.3\times 10^{-4},0.18,+0.88,0.69,-0.51,0.014)\,, \nonumber
	\end{array}
	\right.
	\end{eqnarray}
    respectively give $(\chi^2,\Delta a_\mu)=(13.9,193 \times 10^{-11})$ and %$(\chi^2,\Delta a_\mu)=(14.08,193 \times 10^{-11})$ and
    $(\chi^2,\Delta a_\mu)=(13.7, 201 \times 10^{-11})$. % $(\chi^2,\Delta a_\mu)=(14.10,197 \times 10^{-11})$. 
    As a result, this hybrid scenario, 
    $U_1$ leptoquark in conjunction with $U(1)_{B_3-L_2}$ gauge boson, 
    explains the $B$-physics and $(g-2)_\mu$ anomalous within $2\sigma$.
    }
	
	\bigskip

	\section{Summary}
	
	The recent observational anomalies lead us to consider a vector leptoquark whose couplings with both left and right chiral fermions are essential.  
It affects various channels of $B$-meson decays and generate lepton flavor universality breaking. At the same time, the leptoquark can contribute to $(g-2)_\mu$ with significant enhancement.
We perform the global analysis to several low-energy observables and 
encounter both left- and right-handed
$U_1$ leptoquark couplings. 
Unfortunately, we have not found common parameter region for $B$ and $(g-2)_\mu$ anomalies without additional muon-philic vector boson. 
\scp{Motivate\yjk{d by} this, 
we found the $U_1$ leptoquark in conjunction 
with additional $U(1)_{B_3-L_2}$ gauge boson is able to reconcile the $B$-physics and $(g-2)_\mu$ anomalies within $2\sigma$.
We expect the experimental measurements will be much more improved in the future and the
leptoquark and the new gauge boson will be better tested accordingly. }

	\bigskip
	%\newpage
	
	\section*{Acknowledgments}
	The work is supported by the National Research Foundation of Korea NRF-2021R1A4A2001897, NRF-2019R1A2C1089334 (SCP), and NRF-2020R1I1A1A01066413 (PYT). We thank Andreas Crivellin for discussion on the bottom quark mass in the computation of Muon $(g-2)$. SCP is thankful to Michael Peskin for inviting to the symposium on the muon g-2 anomaly and discussion.  SCP and YJK also acknowledges the support from Yonsei U. and KIAS during his sabbatical period. We acknowledge the hospitality at APCTP where part of this work was done.
	%

	%\newpage
	
	\bibliographystyle{JHEP}
	\bibliography{refs}

\end{document}